\RequirePackage{fix-cm}
\documentclass[smallextended]{svjour3}       
\smartqed  
\usepackage{graphicx}
\usepackage{amssymb,amsmath}
 \journalname{Engineering with Computers}
\begin{document}

\title{Optimized shooting method for finding periodic orbits of
  nonlinear dynamical systems }

\titlerunning{Optimized shooting method for finding periodic orbits}

\authorrunning{W. Dednam \and A. E. Botha } \author{W. Dednam \and
  A. E. Botha }

\institute{W. Dednam \at Departamento de Fisica Aplicada
  \\ Universidad de Alicante, \\ San Vicente del Raspeig, E-03690
  Alicante \\ Spain \and A. E. Botha  \at Department of Physics,
  Science Campus \\ University of South Africa, Private Bag X6
  \\  Florida 1710, South Africa \\ \email{bothaae@unisa.ac.za}  }

\date{Received: date / Accepted: date}

\maketitle

\begin{abstract}

An alternative numerical method is developed to find stable and
unstable periodic orbits of nonlinear dynamical systems. The method
exploits the high-efficiency of the Levenberg-Marquardt algorithm for
medium-sized problems and has the additional advantage of being
relatively simple to implement. It is also applicable to both
autonomous and non-autonomous systems. As an example of its use, it is
employed to find periodic orbits in the R\"{o}ssler system, a coupled
R\"{o}ssler system, as well as an eight-dimensional model of a
flexible rotor-bearing; problems which have been treated previously
via two related methods. The results agree with the previous methods
and are seen to be more accurate in some cases. A simple
implementation of the method, written in the Python programming
language, is provided as an Appendix.

\keywords{Finding periodic orbits \and Levenberg-Marquardt algorithm
  \and Least-squares estimation of nonlinear parameters  \and
  R\"{o}ssler system \and Flexible rotor-bearing system}
\PACS{02.70.-c \and 02.60.Lj \and 02.60.Pn \and 05.45.-a \and
  05.45.Pq} \subclass{ 34B15 \and 34C25 \and 35B10 \and 49N20 \and
  70K42 }
\end{abstract}

\section{Introduction}
In nonlinear dynamical systems, knowledge of the periodic orbits and
their stability is a key aspect of understanding the dynamics. In
particular, the unstable periodic orbits of a chaotic attractor can
offer valuable qualitative as well as quantitative information about
it. The attractor's chaotic trajectories in state space can be
understood intuitively by visualizing the chaotic trajectory as
resulting from a continuous repulsion away from the unstable periodic
orbits that are embedded within the basin of attraction \cite{gal01}.
The relatively recent realization that chaotic trajectories can be
viewed in this manner has led to the development of the \emph{method
  of close returns}, which can be used to extract quantitative
information, such as the Lyapunov exponents, from a time series of the
state space variables \cite{hil00}. In the field of quantitative
finance, for example, such chaotic behavior is finding its way into
the dynamics of markets \cite{fra00}, especially in the periods of
financial crisis.

The problem of finding periodic orbits is essentially a boundary value
problem and there are thus only a few distinct algorithms available
for its solution.  Guckenheimer and Meloon \cite{guc00} have
classified the available methods into three categories: numerical
integration, shooting, and global methods.  Numerical integration
methods have limited application since they are suitable for problems
where the integration of an initial value within the domain of
attraction of a stable periodic orbit converges to the orbit.
Shooting methods compute approximate trajectory segments with an
initial value solver, matching the ends of these trajectory segments
with each other and the boundary conditions, usually by using a root
finding algorithm \cite{deu80}.  Global methods project the
differential equations onto a finite dimensional space of curves that
satisfy the boundary conditions \cite{asc88}.

In recent years collocation methods have become the predominant global
method for finding periodic orbits. Zhou \emph{et al}.~\cite{zho01},
for example, have used the properties of the shifted  Chebyshev
polynomials to transform both autonomous and non-autonomous nonlinear
differential equations into linear and nonlinear algebraic systems,
respectively.  This approach  facilitates the use of algebraic methods
to obtain the periodic solutions of the systems. Their method was
successfully tested on two related  autonomous systems; namely, the
three-dimensional R\"{o}ssler system (RS) \cite{ros76} and
six-dimensional  coupled R\"{o}ssler system \cite{ros96}. However,
obtaining the periods of non-autonomous  systems via their method
proved to be more difficult, since many additional calculations were
required in order to solve the nonlinear algebraic equations that
resulted from the Chebyshev transformation.

More recently, Li and Xu~\cite{li005} developed a generalization of
the shooting method in which the periodic solution and the period of
the system could be found simultaneously. By expanding the residual
function in a Taylor series near the initial condition, the
integration increment could be obtained from an initial value problem
of a set of ordinary differential equations. A comparison of the
method with the work of Zhou \emph{et al}.~\cite{zho01}, was made by
using the RS. Li and Xu~\cite {li005} also successfully applied their
generalized shooting method to a high-dimensional non-autonomous
forced nonlinear system: the model of a Jeffcott flexible
rotor-bearing \cite{cho13,sha90}.

In this article we present an alternative shooting method for finding
the periodic solutions and associated periods of nonlinear
systems. Our method can be viewed as an extension of the generalized
shooting method developed by Li and Xu~\cite{li005}, in that it also
incorporates the period into the system equations. However, instead of
finding the minimum of the residual function by reducing the problem
to that of solving a large set of ordinary differential equations, we
instead apply Levenberg-Marquardt optimization (LMO)
\cite{lev44,mar63,fan12} to obtain the nonlinear parameters which
satisfy the periodic boundary condition. To the best of our knowledge,
such an application of LMO has not been made before.

Lately there have been several modified versions of the  basic
Levenberg-Marquardt algorithm (LMA), and these are finding new
applications in a wide variety of fields. (See, for example,
Refs.~\cite{fan12,fan13}, and the references therein.) In the present
work, however, we make use of the  original algorithm, without
exploring how the performance of our method could benefit from recent
advances in LMO. Such an investigation would only be important for
large-sized problems (i.e. those with a few thousand
weights~\cite{row13})  and is thus beyond the scope of the present
article.

In general the performance comparison of different methods for finding
solutions is not straight-forward.  For example, it is not sufficient
merely to compare different methods for a small number of systems,
since it may be possible to choose a few systems for which one
specific method may perform particularly well. Even if one focuses on
two or three systems, the overall performance of the different methods
should be judged from various perspectives, not only in terms of
computational effort and memory usage, as was traditionally the
norm. In the present work we have therefore not made a detailed
comparison of CPU times and memory usage. All of the examples we have
selected can run comfortably within a few seconds  on an ordinary
desktop computer. Of course the problem of finding periodic solutions
is also highly relevant to much larger dynamical systems, such as
those that occur in continuum models of the human brain
\cite{kim07}. In such models  the un-simplified calculations could
take weeks or even months to complete,  and then the issue of
efficiency does become important. 

The material in this article is organized as follows. In Sec. 2 we
describe the new method, which we call the optimized shooting
method. In Sec. 3 the method is used to find stable periodic orbits of
the RS \cite{ros76}, a problem which was considered  in
Refs.~\cite{zho01,li005}. In Sec. 4 the method is employed to find
unstable periodic orbits of the RS. Several examples of unstable
periodic orbits are discussed. In Sec. 5 we show how the method may be
used to design periodic orbits with specific characteristics, which is
one feature of our method that has the potential for real
engineering-type applications. In Sec. 6 we apply our method to a
six-dimensional (symmetrically) coupled RS and compare the results to
a related method. In Sec.  7 a non-autonomous, 8-dimensional system is
optimized and compared.  The article concludes with a summary of the
main results in Sec. 8. A simple computer implementation of the
method, written in the Python programming language, is provided as an
Appendix.

\section{Optimized shooting method}
Since our aim is to develop a new method for finding periodic orbits
via Levenberg-Marquardt optimization\cite{lev44,mar63} (LMO), we begin
with a  brief description of the Levenberg-Marquardt algorithm (LMA).  

Over the years LMO  has become a standard tool for solving nonlinear
optimization problems in a wide variety of fields. The popularity of
the method stems from the fact that it significantly outperforms
gradient descent and conjugate gradient methods in the optimization of
medium  sized nonlinear models.\cite{row13} Consider the problem of
fitting a function $\tilde{y}=g\left( \tilde{x} ,\boldsymbol{\alpha}
\right) $ to a set of $m$ given data points $(x_{k},y_{k})$, $
k=1,2,\ldots ,m$. Here $\tilde{x}$ is an independent variable and $
\boldsymbol{\alpha }=\left( \alpha _{1},\alpha _{2},\ldots \alpha
_{n}\right) $ is a vector of the system parameters, with $n<m$. To
solve this problem it is convenient to minimize the sum of the
weighted squares of the errors (or weighted residuals) between the
measured data and the fitted function, i.e.  to minimize the quantity
\cite{gav11}
\begin{eqnarray}
\chi ^{2}\left( \boldsymbol{\alpha }\right)  &=&\frac{1}{2}
\sum_{k=1}^{m}\left( \frac{y\left( x_{k}\right) -\tilde{y}\left(
  x_{k},\boldsymbol{\alpha } \right) }{w_{k}}\right)^2 \nonumber
\\ &=&\frac{1}{2}\mathbf{y}^{T}\mathbf{Wy}-\mathbf{y}^{T}
\mathbf{W\tilde{y}}+\frac{1}{2}\mathbf{\tilde{y}}^{T}\mathbf{W
  \tilde{y}} \mbox{.}\label{eq1}
\end{eqnarray}
In Eq.~(\ref{eq1}) the $m \times m$ weighting matrix $\mathbf{W}$ is
diagonal, with $W_{kk}=1/w_{k}^{2}$. Traditionally there are two
methods to obtain the minimum: the gradient descent (or steepest
descent) and Gauss-Newton methods.\cite{gil11} According to the
gradient descent method the perturbation $\mathbf{h}_{gd}$ that moves
the parameters in the direction of steepest descent towards the
minimum is given by
\begin{equation}
\mathbf{h}_{gd}=\beta \mathbf{J}^{T}\mathbf{W}\left( \mathbf{
  y}-\mathbf{\tilde{y}}\right) \mbox{,} \label{eq2}
\end{equation}
where $\mathbf{J}$ is the Jacobian matrix of the function $g$,  and
$\beta $ is a positive scalar that determines the the length of the
step. According to the Gauss-Newton method the required perturbation
is given  by $\mathbf{h}_{gn}$, where
\begin{equation}
\left[ \mathbf{J}^{T}\mathbf{WJ}\right] \mathbf{h}_{gn}=
\mathbf{J}^{T}\mathbf{W}\left( \mathbf{y}-\mathbf{\tilde{y}} \right)
\mbox{.}
\label{eq3}
\end{equation}
The LMA adaptively varies the parameter updates between the gradient
descent and Gauss-Newton update, i.e.
\begin{equation}
\left[ \mathbf{J}^{T}\mathbf{WJ}+\lambda \mathbf{1}\right]
\mathbf{h}_{lm}=\mathbf{J}^{T}\mathbf{W}\left( \mathbf{y}-
\mathbf{\tilde{y}}\right) \mbox{,} \label{eq4}
\end{equation}
where small values of the algorithmic parameter $\lambda$ result in a
Gauss-Newton update and large values of $\lambda $ result in a
gradient descent update.  At a large distance from the function
minimum, the gradient descent method is utilized to provide steady and
convergent progress towards the solution.  As the solution approaches
the minimum, $\lambda $ is adaptively decreased and the LMA approaches
the Gauss-Newton method, for which the solution typically converges
more rapidly to the local minimum. The update relationship suggested
by Marquardt \cite{mar63} is given by 
\begin{equation}
\left[ \mathbf{J}^{T}\mathbf{WJ}+\lambda \mathrm{diag}\left(
  \mathbf{J}^{T}\mathbf{WJ}\right) \right] \mathbf{h}_{lm}=
\mathbf{J}^{T}\mathbf{W}\left( \mathbf{y}-\mathbf{\tilde{y}} \right)
\mbox{.} \label{eq5}
\end{equation}
In this work we make use of the Python function \verb|leastsq()|,
which provides an efficient implementation of the LMA, to minimize the
residual $\mathbf{y}-\mathbf{\tilde{y}}$. For simplicity, in the
following sections, we will denote the residual simply as $\mathbf{R}
\equiv  \mathbf{y}- \mathbf{\tilde y}$    

The optimized shooting method developed here is applicable to any
dynamical system that can be written in the standard form 
\begin{equation}
\frac{\mathrm{d}\mathbf{x}}{\mathrm{d}t}={\mathbf{f}}\left( \mathbf{x}
,\boldsymbol{\alpha} ,t\right) \mbox{.}  \label{eq6}
\end{equation}
In Eq.~(\ref{eq6}) the functions ${\mathbf{f}}=\left(
f_{1},f_{2},\ldots ,f_{N}\right) ^{T}$ are functions of the dynamical
variables $\mathbf{x=} \left( x_{1},x_{2},\ldots ,x_{N}\right) $, as
well as the system parameters $ \boldsymbol{\alpha} $, and time $t$. A
periodic solution of Eq.~(\ref{eq6}) is a closed trajectory for which
there exists a positive real number $T$, such that $ \mathbf{x}\left(
T\right) =\mathbf{x}\left( 0\right) $. The quantity $T$ is called the
period of solution. Since we are interested in finding the periodic
solutions to Eq.~(\ref{eq6}), we will rewrite Eq.~(\ref{eq6}) in terms
of a dimensionless time $\tau $, such that $t=T\tau $. This
substitution  produces an equivalent equation, given by 
\begin{equation}
\frac{\mathrm{d}\mathbf{x}}{\mathrm{d}\tau }=T{\mathbf{f}}\left(
\mathbf{x} ,\boldsymbol{\alpha} ,T\tau \right) \mbox{.}  \label{eq7}
\end{equation}
Since $\tau $ is measured in units of the period $T$, Eq.~(\ref{eq7})
has the advantage that the boundary condition for a periodic solution
can now be expressed as $\mathbf{x}\left( \tau = 0 \right)
=\mathbf{x}\left( \tau = 1 \right)  $. Starting from an initial
condition  $\mathbf{x}\left(\tau = 0 \right)$ one can thus integrate
the  equation numerically over exactly one period by letting $\tau $
run from zero to one. This integration then allows one to define the
residual, which in the above notation, can be written formally as
\begin{equation}
\mathbf{R} = T\int_{0}^{1} \mathbf{f}\left(
\mathbf{x},\boldsymbol{\alpha},T\tau \right) \mbox{d}\tau \mbox{.}
\end{equation}

In practice there are many different ways of defining the residual and
these will depend on which  of the quantities (initial conditions and
parameters) and how many  are to be optimized. However, for the
purposes of describing the basic method  we will assume that we wish
to compute a periodic solution passing through some initial point
$\mathbf{x}\left( 0\right) $, at fixed values of the  parameters
$\boldsymbol{\alpha} $. For this case the residual can be written
explicitly as 
\begin{eqnarray}
\mathbf{R}& = &(\mathbf{x}\left( 1\right) -\mathbf{x}\left( 0\right)
,\mathbf{x} \left( 1+\Delta \tau \right) -\mathbf{x}\left( \Delta \tau
\right) , \nonumber \\ & & \, \, \, \, \ldots , \mathbf{x}\left(
1+\left( p-1\right) \Delta \tau \right) -\mathbf{x}\left( \left(
p-1\right) \Delta \tau \right) ) \mbox{.} \label{eq8}
\end{eqnarray}
In Eq.~(\ref{eq8}), $\Delta \tau $ is the numerical integration step
size, $\mathbf{x} \left( 0\right) $ is the sought after initial point,
and $p=1,2,3,\ldots $ is a natural number which should be chosen large
enough to ensure that $\mathbf{R}$ has an equal or greater number of
components than the number of quantities which are to be
optimized. This restriction on the choice of $p$ is a requirement of
the LMA. (See, for example, Ref.~\cite{lan04}.) In view of
Eq.~(\ref{eq8}) it can be seen that the number of components of
$\mathbf{R}$ will in general be given by $pN$, where $N$ is the system
dimension. The strategy now is to use LMO to efficiently minimize the
residual, noting that ${\mathbf{R}}={\mathbf{0}}$ gives periodic
solutions.

To better illustrate the method we next supply a concrete example.
Consider the problem of obtaining a periodic solution for the famous
R\"{o}ssler  system \cite{ros76}. When written in the form of
Eq.~(\ref{eq7}),  R\"{o}ssler's system is given by 
\begin{eqnarray}
\dot{x}_{1} &=&-T\left( x_{2}+x_{3}\right)   \nonumber \\ \dot{x}_{2}
&=&T\left( x_{1}+ax_{2}\right)   \label{eq9} \\ \dot{x}_{3} &=&T\left(
b+x_{3}\left( x_{1}-c\right) \right)   \nonumber
\end{eqnarray}
In order to facilitate a comparison with previous work
\cite{zho01,li005}, we will fix the parameters at the predetermined
values of $a=0.15$, $b=0.2$ and $c=3.5$. For these values it is known
that the system has a stable period-1 orbit, with one point on the
orbit reported by Zhou \emph{et al}.~\cite{zho01} to be
$\left(2.7002161609,3.4723025491,3.0\right)$, with a principal period
$T=5.92030065$. In this case there are four quantities which need to
be optimized: $x_{1}\left( 0\right) $, $ x_{2}\left( 0\right) $,
$x_{3}\left( 0\right) $ and $T$. The smallest possible choice of $p$
is therefore $p=2$, giving a residual with $6$ components, i.e. 
\begin{eqnarray}
\mathbf{R} &=&(x_{1}\left( 1\right) -x_{1}\left( 0\right) ,x_{2}\left(
1\right) -x_{2}\left( 0\right) ,x_{3}\left( 1\right) -x_{3}\left(
0\right) , x_{1}\left( 1+\Delta \tau \right) \nonumber \\ & & \, \, \,
-x_{1}\left( \Delta \tau \right) ,  x_{2}\left( 1+\Delta \tau \right)
-x_{2}\left( \Delta \tau \right) , x_{3}\left( 1+\Delta \tau \right)
-x_{3}\left( \Delta \tau \right) ) \mbox{.}  \label{eq10}
\end{eqnarray}
We note here that the residual is a function of all four optimization
parameters, as well as the system parameters $a$, $b$ and $c$; since
it depends on these implicitly through Eq.~(\ref{eq9}). Rather than
locating any initial point on the orbit we will try to locate a point
for which $ x_{3}\left( 0\right) =3.0$. This will facilitate a direct
comparison with the point that was located by Zhou \emph{et
  al}.~\cite{zho01}. We therefore exclude $x_{3}$ from the
minimization  process. To clarify how this is done in practice, we
have provided an Appendix containing a complete Python implementation
of the above example, with additional comments written inside the
code. Note that in this example code the function \verb|leastsq()|
uses finite differences to approximate the Jacobian matrix. This
method of determining the Jacobian is known to be more costly and may
in fact be impossible for very stiff systems. However, in such cases
the efficiency of the code may be improved by specifying the
functional form of the Jacobian matrix explicitly as  one of the
optional input arguments to the function \verb|leastsq()| (this is not
done in the example code).  The results from the code will be
discussed in the next section.

\section{Comparison to two closely related methods}
The results obtained from the code in Appendix A are highly accurate
in comparison to those obtained from either the collocation method, by
Zhou  \emph{et al}.~\cite{zho01}, or the generalized shooting method,
by Li and Xu~\cite{li005}. Table~\ref{tab1} lists the initial
conditions and periods for comparison.

\begin{figure}[htp!]
\centering \includegraphics[width=0.9\textwidth]{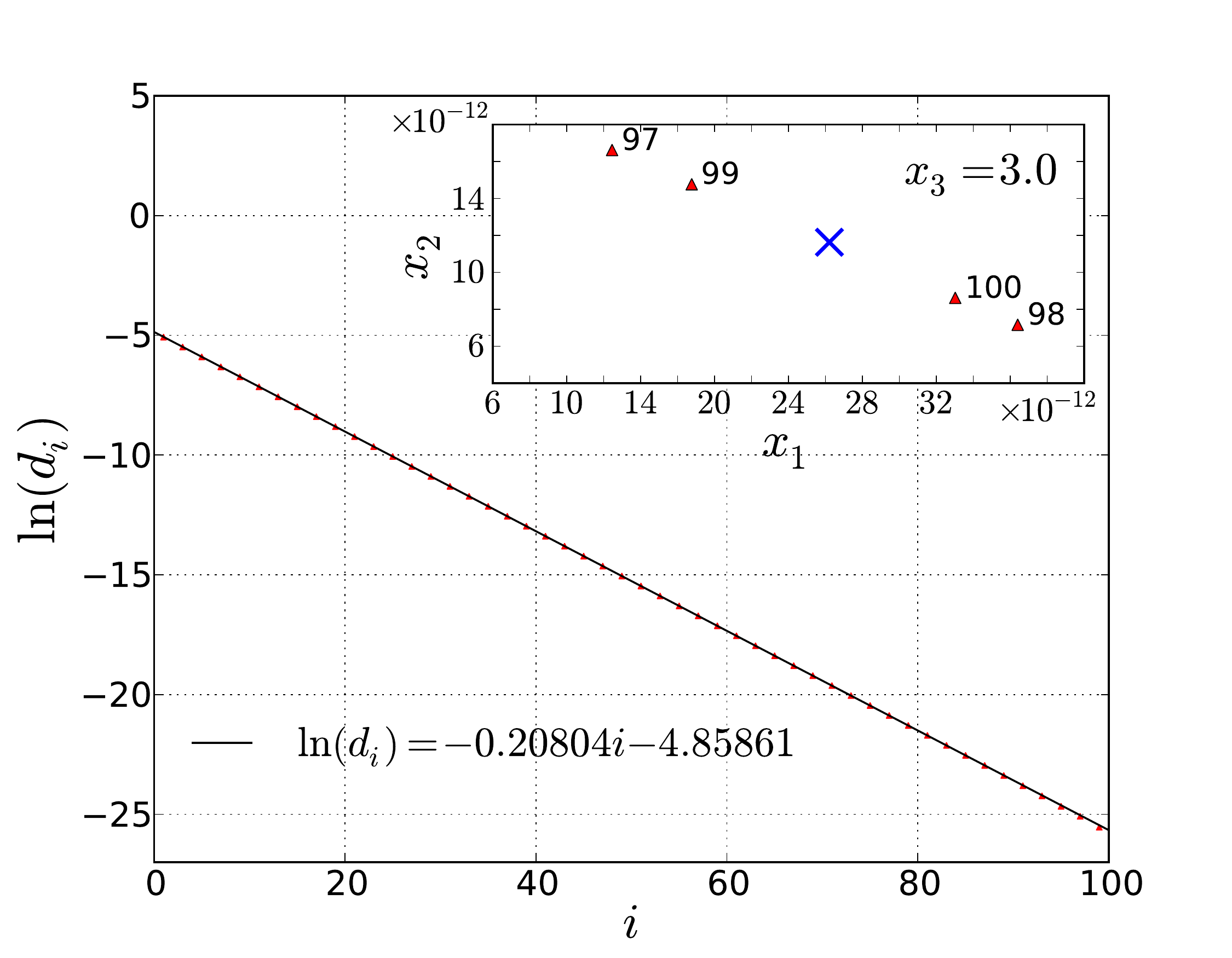}
\caption{(Color online) Logarithmic plot of distance to convergence
  versus number of iterations, from the point on the orbit found by
  Zhou \emph{et al}.~\cite{zho01}, to the more accurate point found by
  the code in the Appendix after only 20 additional integration
  cycles.  More than  $100$  ordinary integration cycles are required
  before the \emph{Zhou et al}. orbit  converges to a distance smaller
  than $6\times 10^{-12}$. The inset  shows a Poincar\'{e} section
  through the plane $x_{3}=3.0$,  with successive intersections of the
  converging orbit indicated by (red) triangles, and the more accurate
  point by a (blue) cross. The bottom left corner of the inset
  correspond to $x_1=2.628655670296$ and $x_2=3.509456205164$. These
  values should be added to the  corresponding axis labels.}
\label{fig1}
\end{figure}

\begin{table}
\centering
\begin{tabular}{|c||c|c|c|}
\hline & $x_{1}\left( 0\right) $ & $x_{2}\left( 0\right) $ & $T$
\\ \hline\hline $^{a}$ & \multicolumn{1}{|l|}{$2.6286556703142154$} &
\multicolumn{1}{|l|}{$ 3.5094562051716300$} &
\multicolumn{1}{|l|}{$5.920340248194$} \\ \hline $^{b}$ &
\multicolumn{1}{|l|}{$2.7002161609$} & \multicolumn{1}{|l|}{$
  3.4723025491$} & \multicolumn{1}{|l|}{$5.92030065$} \\ \hline $^{c}$
& --- & --- & \multicolumn{1}{|l|}{$5.92190215$} \\ \hline
\end{tabular}

{\footnotesize $^{a}$Present work. $^{b}$Zhou \emph{et
    al}.~\cite{zho01}.  $^{c}$Li and Xu~\cite{li005} -- no point on
  the orbit was provided.}
\caption{Optimized initial coordinates $x_1(0)$, $x_2(0)$ and period
  $T$ of the period-1 solution to Eq.~(\ref{eq9}), obtained for the
  predetermined values of control parameters, with $x_{3}\left( 0
  \right) =3.0$.} 
\label{tab1}
\end{table}
In the case of the collocation method, the fourth column in
Table~\ref{tab1}  shows that the periods agree only to four decimal
places, while for the generalized shooting method, the agreement is
even worse (only two decimal places). Since the maximum error in the
residual is $2.1\times 10^{-14}$, we estimate our calculated period to
be accurate to $12$ decimal places, as indicated in
Table~\ref{tab1}. The discrepancy between our result and the other two
methods can be understood by examining the convergence of the
trajectory towards the closed limit cycle. To this end we have
integrated the orbit found by Zhou \emph{et al}.~\cite{zho01} for
$100$ periods, starting from the initial condition
$x_{10}=2.7002161609$, $x_{20}=3.4723025491$, as listed in the second
row of Table~\ref{tab1}. In Fig.~\ref{fig1} we have plotted the distances,
$d_{i}=\sqrt{\left( x_{1i}-x_{1}\left( 0\right) \right) ^{2}+\left(
  x_{2i}-x_{2}\left( 0\right) \right) ^{2}}$ against the number of
additional integration cycles (each cycle is one period long),  where
$\left( x_{1}\left(0\right),x_{2}\left( 0\right) \right) $ is the
point on the orbit obtained via the optimized shooting method (listed
in the first row of Table 1). The inset in Fig.~\ref{fig1} shows a
Poincar\'{e} section through the plane $x_{3}=3.0$, where the 
point on the orbit, found by the optimized shooting method, has been
plotted as a blue cross. The triangular markers in the inset show the
successive crossing of the other orbit that slowly converges toward
the  cross. Note that labels for the horizontal and vertical axes of
the inset  should be multiplied by $10^{-12}$ and then added to
$x_1=2.628655670296$ and $x_2=3.509456205164$, respectively. The last
two coordinates are those of the bottom left corner of the inset.

As may be expected, the distances $d_{i}$ converge exponentially
toward the stable limit cycle. The decay constant was found to be
$-0.20804$. From the Poincar\'{e} section we see that it takes
approximately an additional $ 100$ cycles before the orbit found by
Zhou \emph{et al}.~\cite{zho01} converges to a distance of less than
$6\times 10^{-12}$ away from the orbit found by the optimized shooting
method. The same problem, namely the incomplete convergence of the
orbit, is also responsible for inaccuracies in the results reported by
Li and Xu~\cite{li005} and it points to an important advantage of the
present method. In the case of the code listed in Appendix A,
convergence to the final orbit is complete after only $20$ integration
cycles, as opposed to the more than $100$ additional cycles that are
required before the other two methods converge to  the true periodic
orbit to the same accuracy. Moreover, in many  applications one is
interested in obtaining unstable periodic orbits, but numerical
integration methods will fail to converge to unstable orbits.  As we
shall see in the next section, the optimized shooting method is also
suitable for finding unstable periodic orbits.

To conclude this section we note that we have also tested and compared
the new method for the case of period-2 and higher orbits. Figure 2
summarizes  our results for the RS, by showing the phase portraits of
the obtained period-1 and period-2 orbits.
\begin{figure}[htp!]
\centering \includegraphics[width=0.49\textwidth]{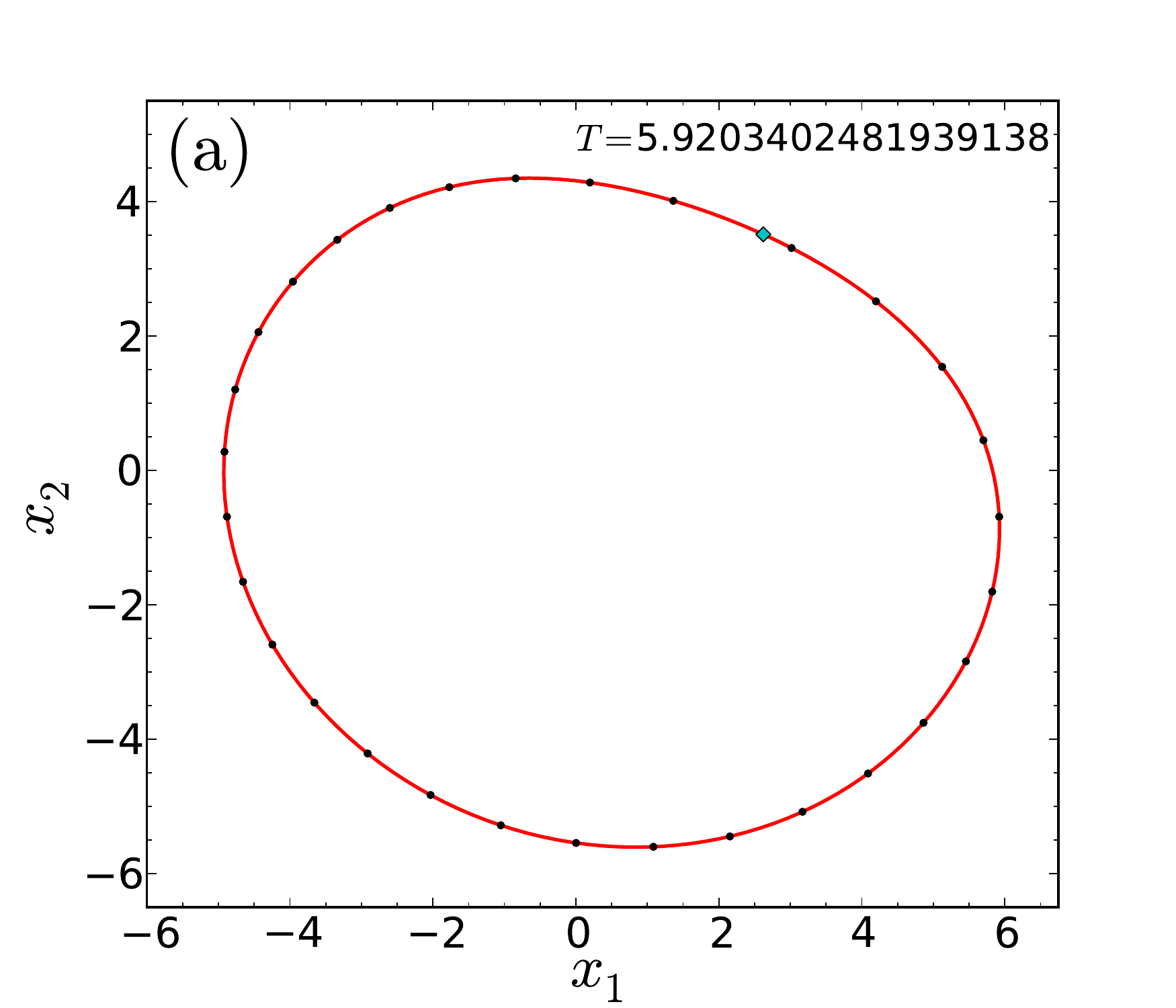}
\includegraphics[width=0.49\textwidth]{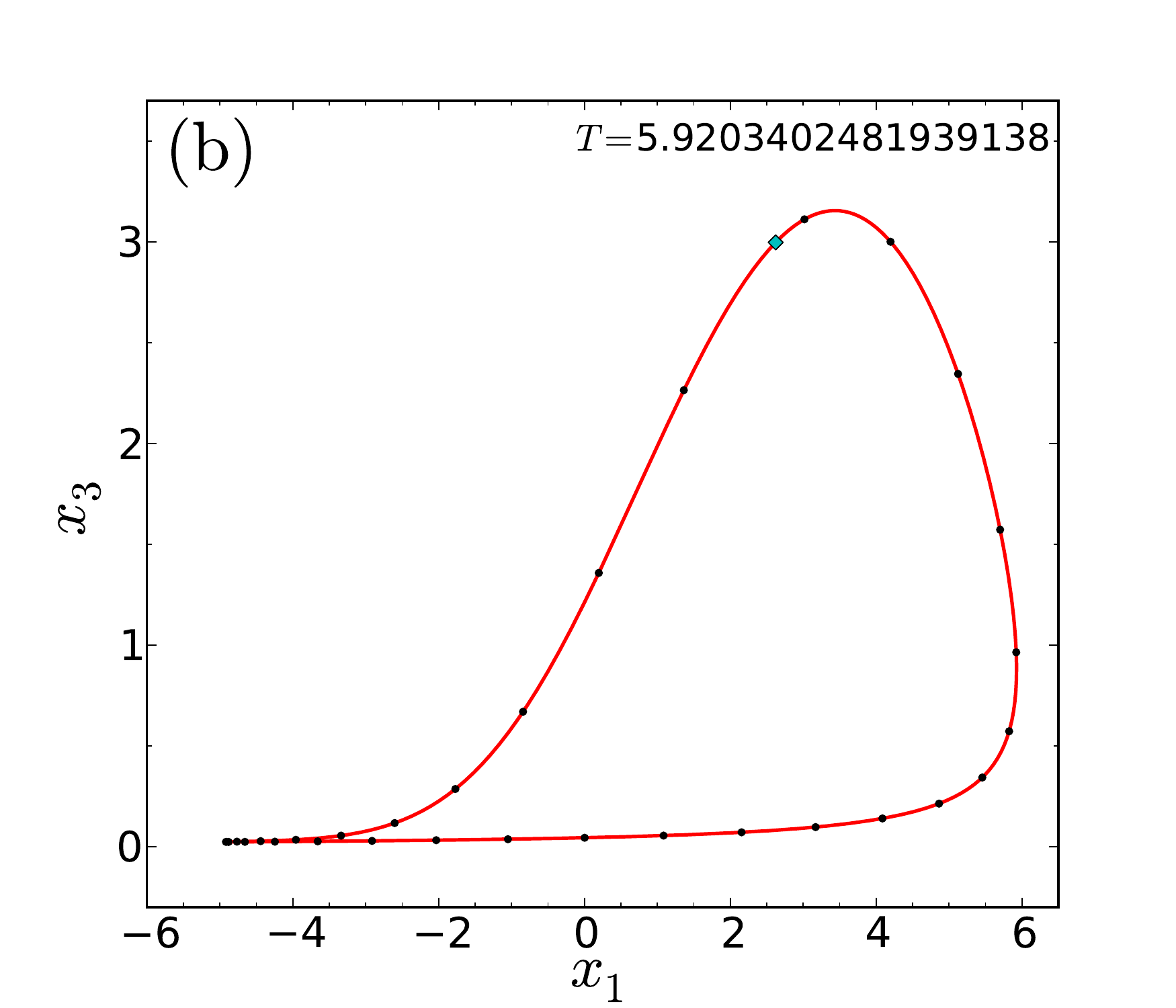}

\includegraphics[width=0.49\textwidth]{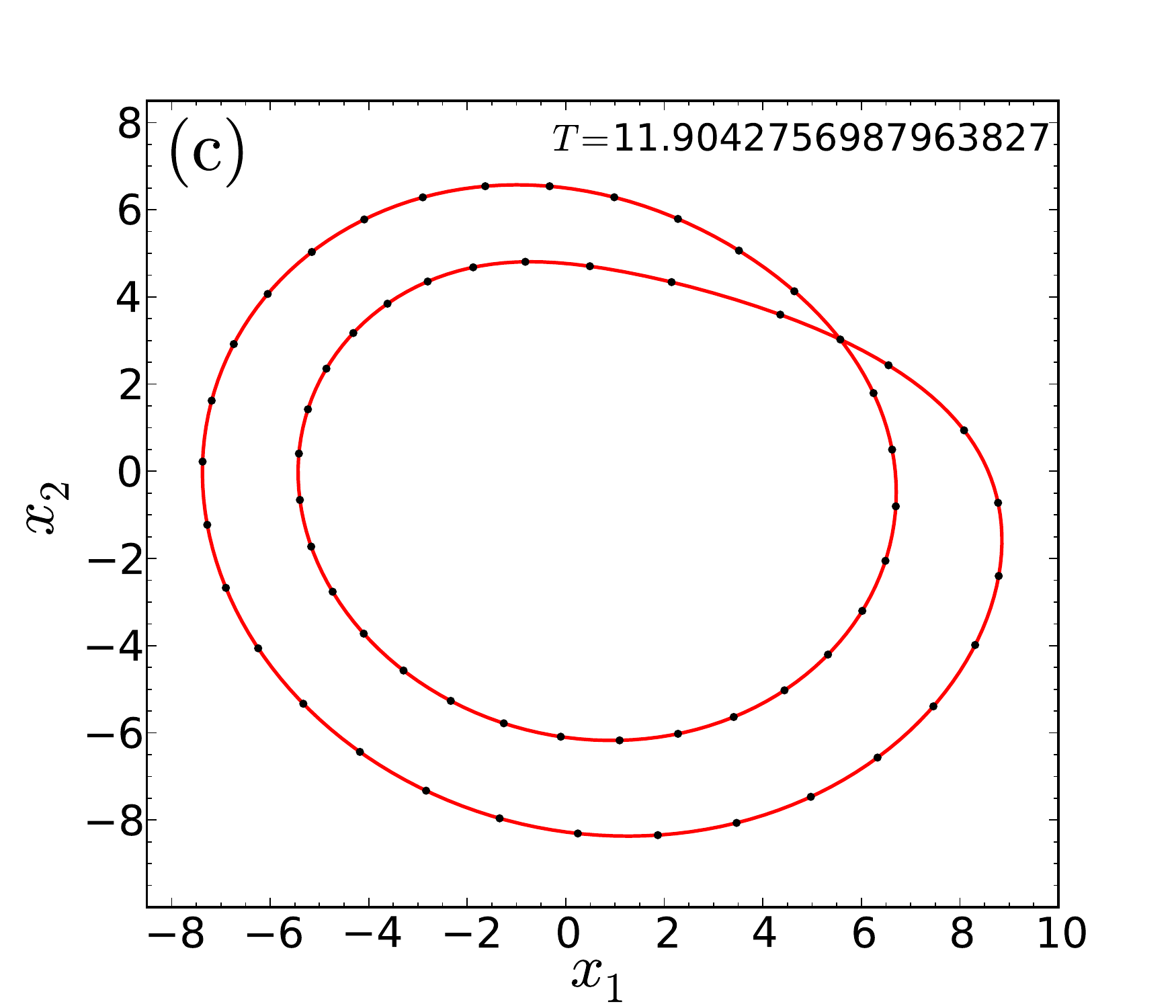}
\includegraphics[width=0.49\textwidth]{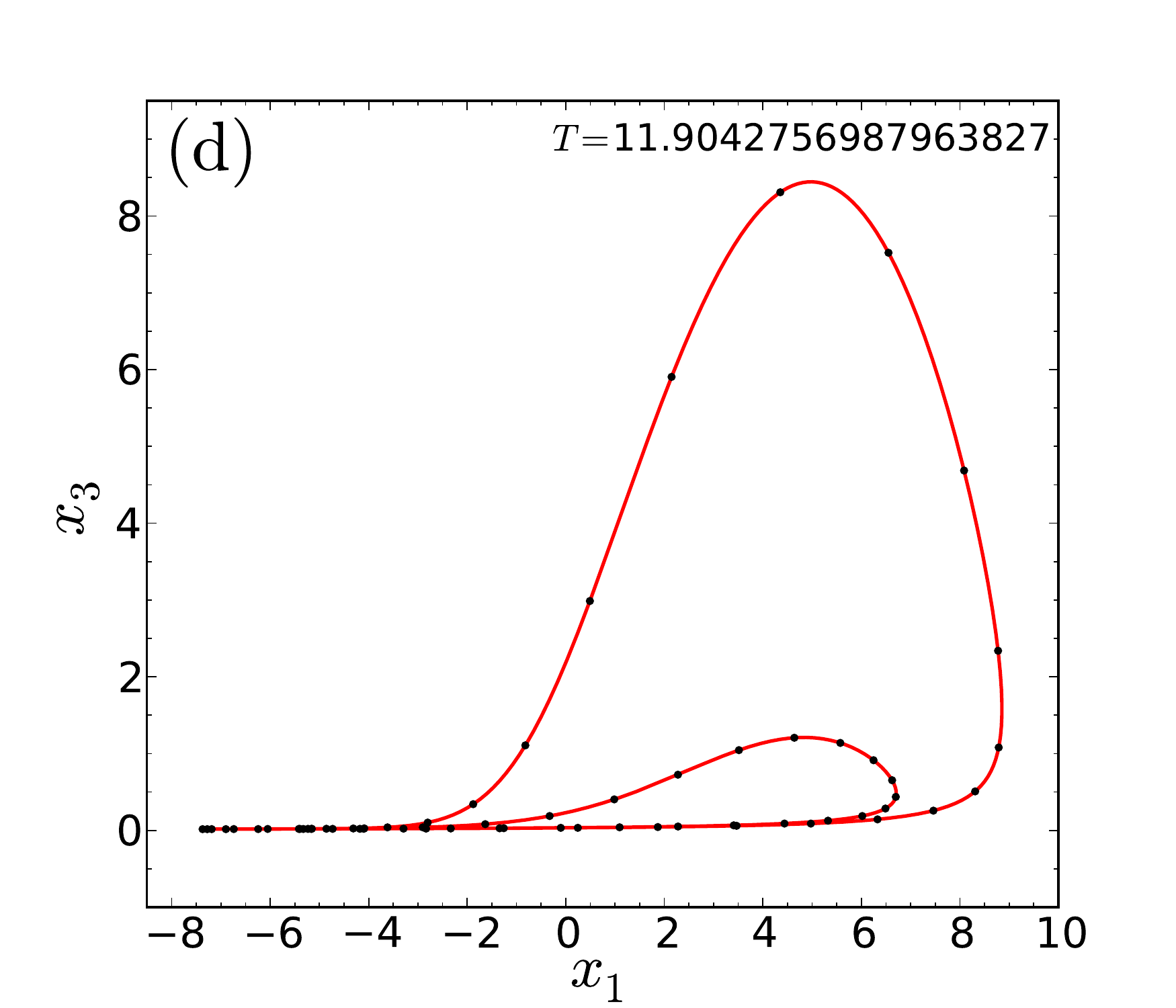}
\caption{(Color online) Comparison of phase portraits obtained via the
  optimized shooting method, with that of Ref.~\cite{li005}. Figures
  (a)  and (b) show phase portraits of the period-1 solution obtained
  by solving  Eq.~(\ref{eq9}) with $c=3.5$. Similarly. Figs. (c) and
  (d) show the  period-2 solution, with $c=5.0$. The optimized periods
  are given in the  upper right-hand corners of each figure. For
  Figs. (a) and (b), the precise coordinates of the point marked by
  (blue) diamond markers  are listed in the first row of
  Table~\ref{tab1}.}
\label{fig2}
\end{figure}
Figures 2(a) to (d) correspond very closely to to Figs. 1 to 4 in
Ref.~\cite{li005}. In the case of Figs. 2(c) and (d), for the
parameter $c=5.0$, Li and Xu calculated the period of the bifurcated
orbit to be $T=11.89692819$.  From our calculated value, shown in
upper right corner of Fig.~\ref{fig2}(d), we see  that the value that was
calculated by Li and Xu~\cite{li005} is again only accurate to two
decimal places.

\section{Unstable periodic orbits in the R\"{o}ssler system} \label{sec4}
When applied to the RS the optimized shooting method finds, in
addition to stable periodic orbits,  a large number of unstable
periodic orbits. In what follows we will characterize the stability of
these orbits by calculating the characteristic (Floquet) multipliers,
i.e. the eigenvalues of the monodromy matrix, according to the method
developed by Lust~\cite {lus01}. The monodromy matrix
$\mathbf{M}\left( t\right) $ is the solution at time $T$ of the
variational equation 
\begin{equation}
\frac{d\mathbf{M}}{dt}=\mathbf{JM}\mbox{, \ \ with }\mathbf{M} \left(
0\right) =\mathbf{1}\mbox{,}  \label{eq11}
\end{equation}
where $\mathbf{J}$ is the system Jacobian. One of the Floquet
multipliers, called the trivial multiplier, is always one. Its
eigenvector is tangential to the limit cycle at the initial point $
\mathbf{x}\left( 0\right) $. A periodic solution is asymptotically
stable if the modulus of each Floquet multiplier, except the trivial
one, is strictly less than one. Otherwise, if one or more of the
multipliers is greater than one in modulus, the solution is
asymptotically unstable. For the stable orbits that were seen in
Fig.~\ref{fig2}, for example, the moduli of the  largest nontrivial multipliers
are found to be $\mu _{\max }=0.812252$ (in the case of
Figs.~\ref{fig2}(a) and (b)) and $\mu _{\max }=0.572052$ (in the case
of Figs.~\ref{fig2}(c) and (d)), respectively.

As a further test of the optimization method we have also optimized
the parameters $a$, $b$ $c$ and $T$ in the R\"{o}ssler system, for
fixed initial values of the coordinates $(x_1,x_2,x_3)$. 

As the initial condition we systematically chose the coordinates to
lie on three-dimensional cubic grids of varying sizes,  starting  from
$8^{3}$ grid points, on a cube surrounding the origin of side length
$1$,  and systematically increasing the number of grid points and the
cube length. In total, literally thousands of periodic orbits were
thus obtained for the R\"{o}ssler system by optimizing the four
parameters. In some cases more than one periodic orbit (corresponding
to different sets of optimized parameters) was found to pass through
the initial condition. In most cases the found orbits were
unstable. After examining several hundreds of these orbits
graphically, we have concluded that they are qualitatively of two
different types, depending on the sign of the $x_{3}$ coordinate.

\begin{figure}[htp!]
\centering \includegraphics[width=0.49\textwidth]{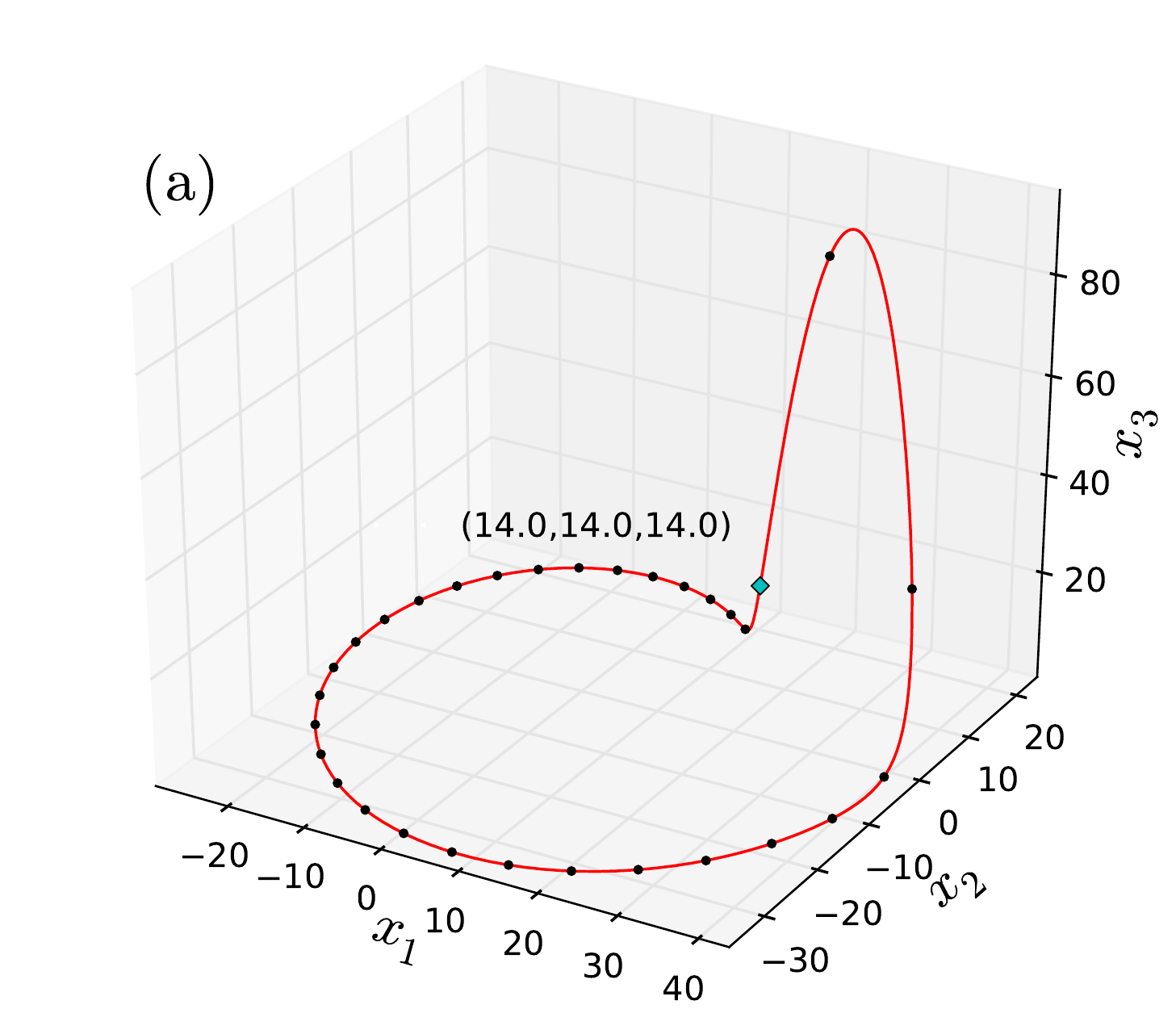}
\includegraphics[width=0.49\textwidth]{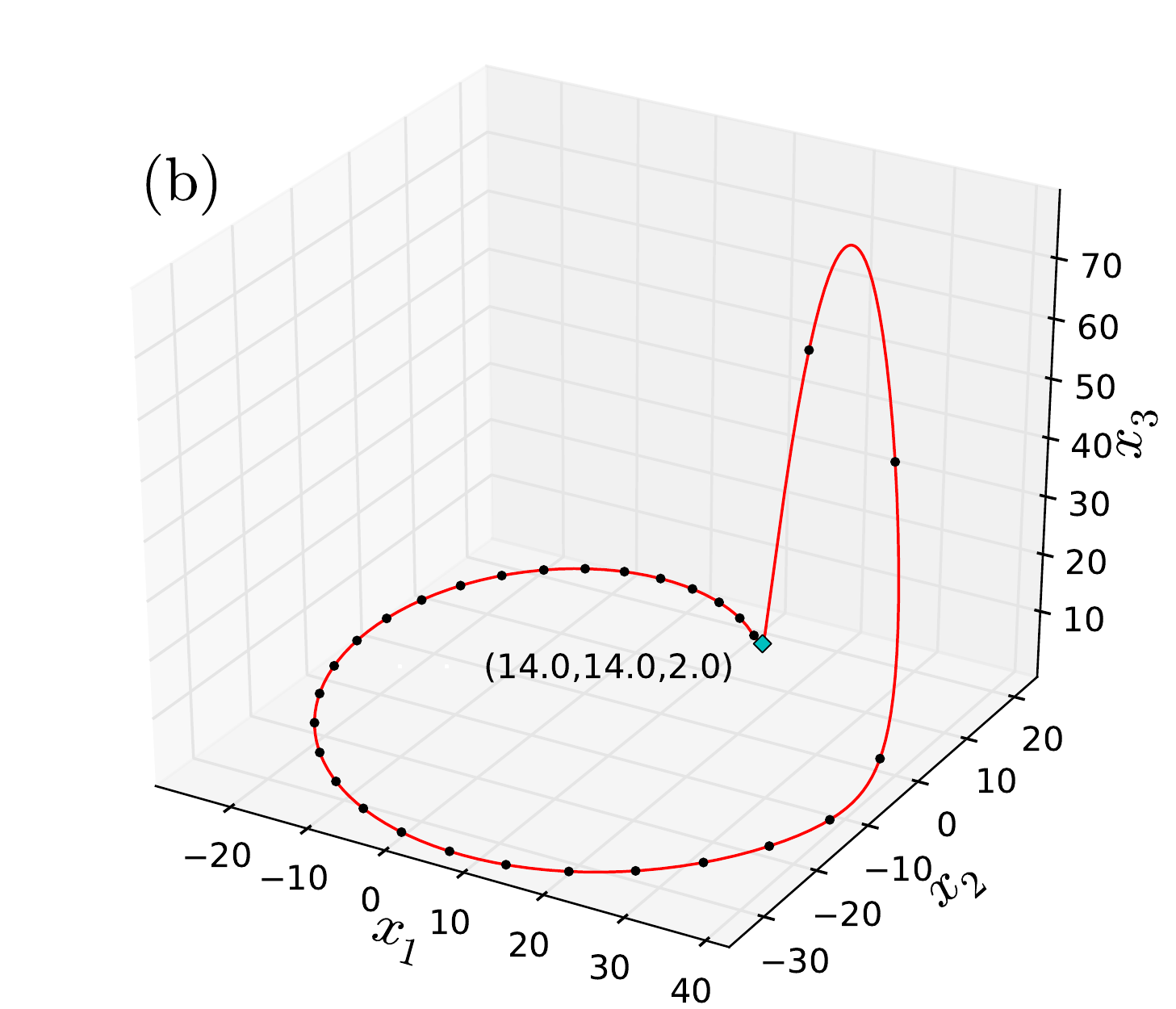}

\includegraphics[width=0.49\textwidth]{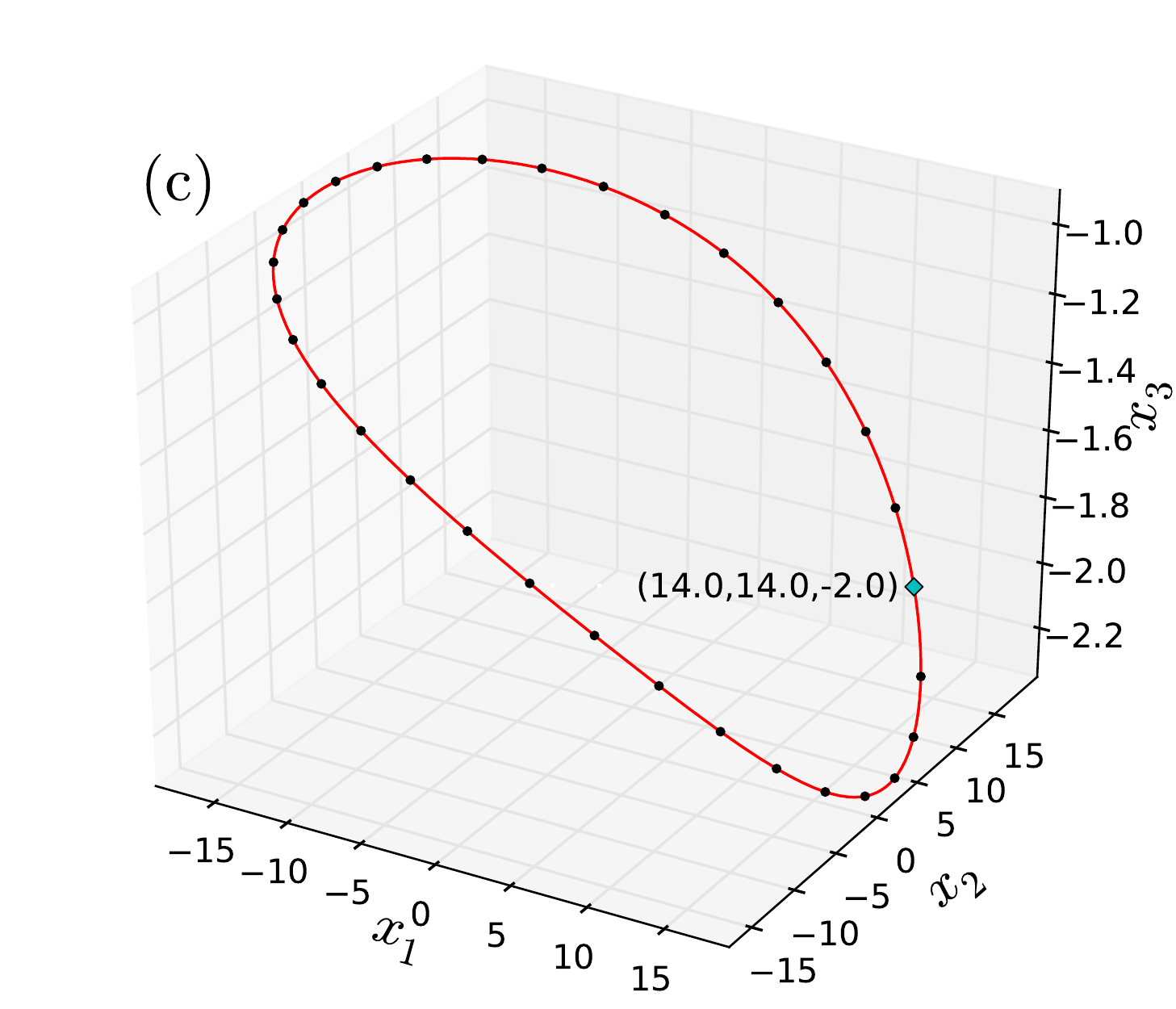}
\includegraphics[width=0.49\textwidth]{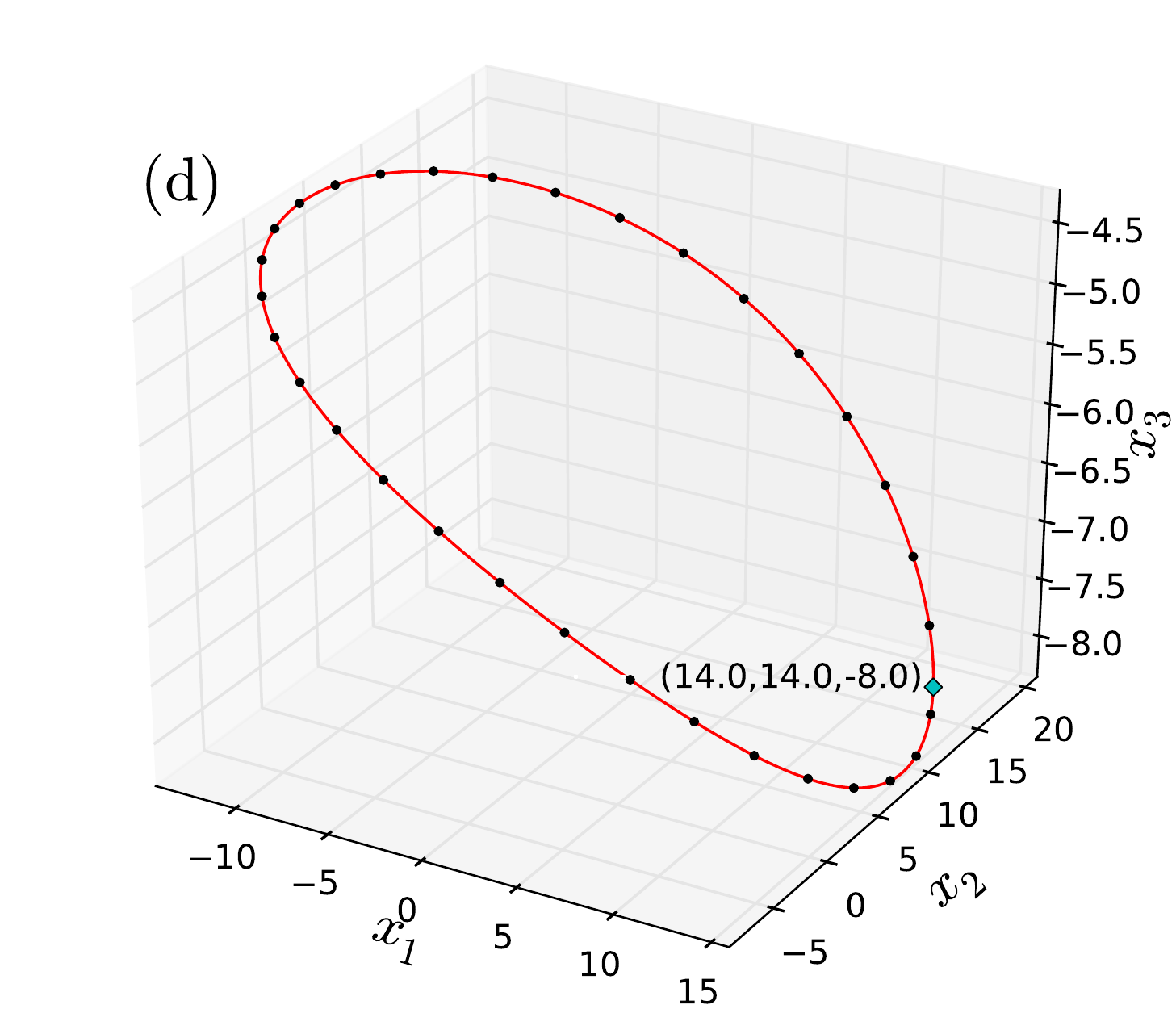}
\caption{(Color online) Qualitative differences between periodic
  orbits of the  R\"{o}ssler system in the half space $x_{3}>0$, shown
  in (a) and (b), and in the half space $x_{3}<0$, shown in (c) and
  (d). Note the characteristic peak of the orbit in the $x_{3}$
  direction, for $x_{3}>0$.}
\label{fig3}
\end{figure}

To illustrate the main qualitative difference between the two types of
orbits,  we show four of the found orbits in Fig.~\ref{fig3}. The orbits in
Figs.~\ref{fig3}(a) and (b) lie entirely in the half space $x_{3}>0$ and they
all have a distinctive peak in the $x_{3}$ coordinate direction. 
This peak is related to an exponential  growth in the $x_{3}$
coordinate, when $x_{1}>0$. In this region the nonlinear term in the
third equation of the RS tends to dominate, causing the exponential
growth. However, depending on the value of the parameter $c$, as well
as the other parameters in the system, the exponential growth in the
coordinate $x_{3}$ eventually changes to exponential decay; either
because the term $-cx_{3}$ becomes significant in comparison to the
nonlinear term or else because of a change in sign of the coordinate
$x_{1}$ (or both). Thus, qualitatively, all the obtained orbits for
which $x_{3}>0$ were roughly elliptical, when projected onto the $
x_{1}x_{2}$-plane, containing a characteristic peak in the $x_{3}$
direction, as show in Figs.~\ref{fig3}(a) and (b). On the other hand,
orbits for which $x_{3}<0$ were also elliptical in the
$x_{1}x_{2}$-plane, but these did not have the exponential peak that
was seen for the $x_{3}>0$ orbits. Two examples of orbits in the lower
half plane  are shown in Figs.~\ref{fig3}(c) and (d). The optimized
parameters for these orbits, together with the modulus of the largest
non-trivial multiplier $\mu_{\max }$ and the magnitude of the located
minimum in the residual $R_{\min }$ are listed in Table~\ref{tab2} for
convenience. Each row in Table~\ref{tab2} is labeled by the
corresponding figure number.

\begin{table}
\begin{tabular}{|rccc|}
\hline Fig. & \multicolumn{1}{|c}{$T$} & \multicolumn{1}{|c}{$R_{\min
  }\left( 10^{-13}\right) $} & \multicolumn{1}{|c|}{$\mu _{\max }$}
\\ \hline\hline 3(a) & \multicolumn{1}{|c}{6.0368126768371511} &
\multicolumn{1}{|c}{6.07} &  \multicolumn{1}{|c|}{5.635982} \\ \hline
3(b) & \multicolumn{1}{|c}{6.1360301575904730} &
\multicolumn{1}{|c}{2.71} &  \multicolumn{1}{|c|}{4.552192} \\ \hline
3(c) & \multicolumn{1}{|c}{6.2906328270213132} &
\multicolumn{1}{|c}{0.08} & \multicolumn{1}{|c|}{1.037877} \\ \hline
3(d) & \multicolumn{1}{|c}{6.3510302249381043} &
\multicolumn{1}{|c}{0.87} &  \multicolumn{1}{|c|}{1.070770} \\ \hline
&  &  &  \\  &  &  &  \\ \hline Fig. & \multicolumn{1}{|c}{$a$} &
\multicolumn{1}{|c}{$b$} &  \multicolumn{1}{|c|}{$c$} \\ \hline\hline
3(a) & \multicolumn{1}{|c}{0.2639519863856384} & \multicolumn{1}{|c}{
  0.0076361389339972} & \multicolumn{1}{|c|}{28.4349476839393454}
\\ \hline 3(b) & \multicolumn{1}{|c}{0.2511744525989691} &
\multicolumn{1}{|c}{ 0.0681976088482620} &
\multicolumn{1}{|c|}{29.4473072446435609} \\ \hline 3(c) &
\multicolumn{1}{|c}{-0.0360057068462920} & \multicolumn{1}{|c}{
  -56.9446489079999978} & \multicolumn{1}{|c|}{42.8808270302575707}
\\ \hline 3(d) & \multicolumn{1}{|c}{-0.1323654753343560} &
\multicolumn{1}{|c}{ -262.2015992062510463} &
\multicolumn{1}{|c|}{46.9462750789147378} \\ \hline
\end{tabular}
\caption{Optimized periods $T$, minimum residuals $R_{\min}$ and the
  moduli of the largest non-trivial Floquet multipliers $\mu_{\max}$
  corresponding to the  the unstable periodic orbits shown in
  Fig.~\ref{fig3}, for the optimized  control parameters $a$, $b$ and
  $c$.}
\label{tab2}
\end{table}

It is easy to see why closed orbits cannot cross the $x_{3}=0$
plane. If such orbits existed, they would have to cross the plane
twice; once with $\dot{x}_3 > 0$ and once with $\dot{x}_3 <
0$. However, as Eq.~(\ref{eq9}) shows, for $x_3=0$ the  sign of
$\dot{x}_3$ is determined by the parameter $b$.  Thus $\dot{x}_3$ can
be either positive or negative, but its sign never alternates for a
given set of parameters. If one does try to construct a periodic orbit
passing through the plane $x_3=0$, the optimized shooting method
simply fails to converge, indicating that the orbit does not exist.

To conclude this section we also investigate the degree of instability
in the calculated unstable periodic orbits. The degree of instability
is particularly important for unstable solutions with long periods
(especially those embedded in strange attractors of a chaotic
system). The latter problem is considered to be hard to solve and is
therefore a good test of the present method.

\begin{figure}[htp!]
\centering \includegraphics[width=0.49\textwidth]{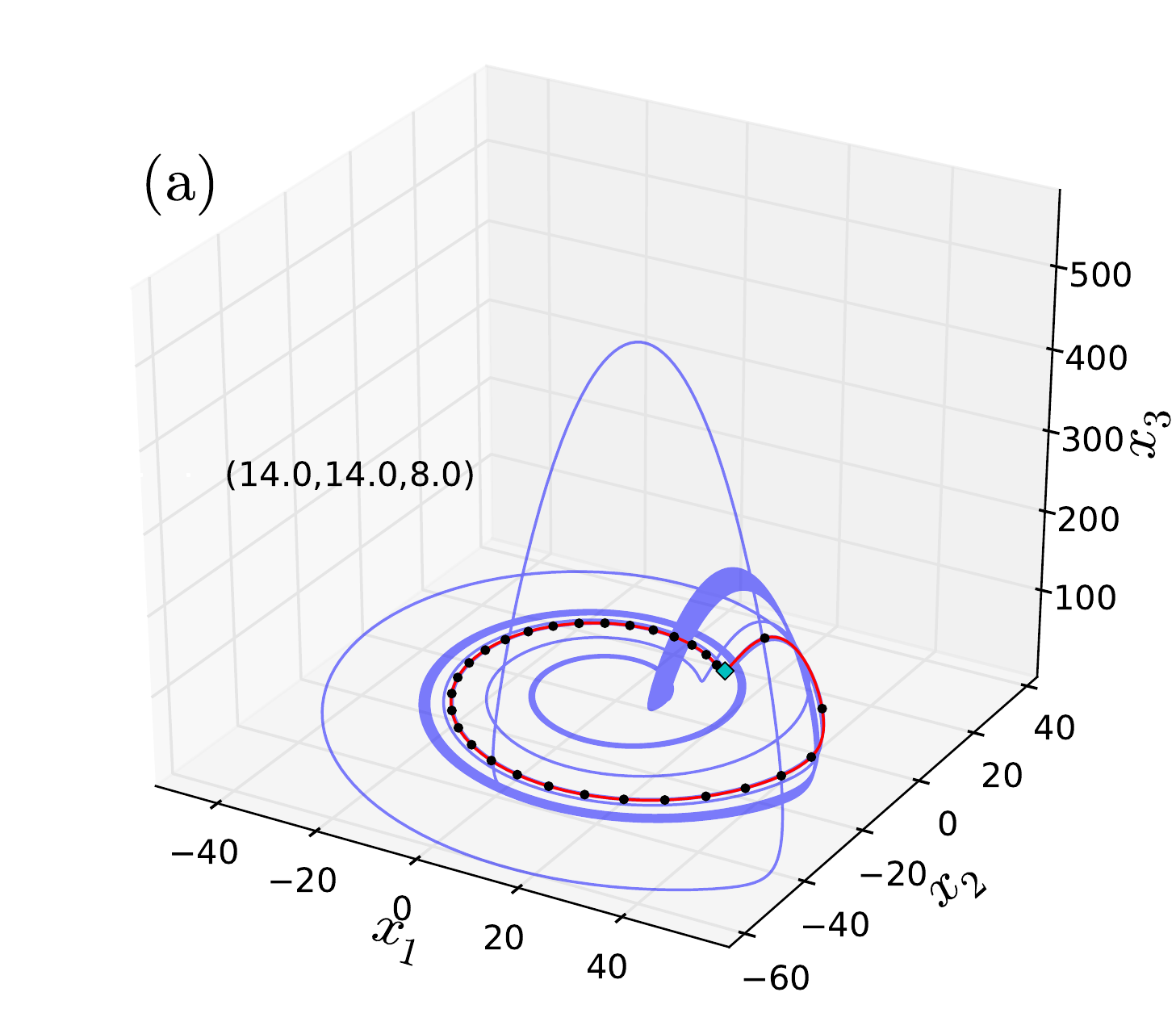}
\includegraphics[width=0.49\textwidth]{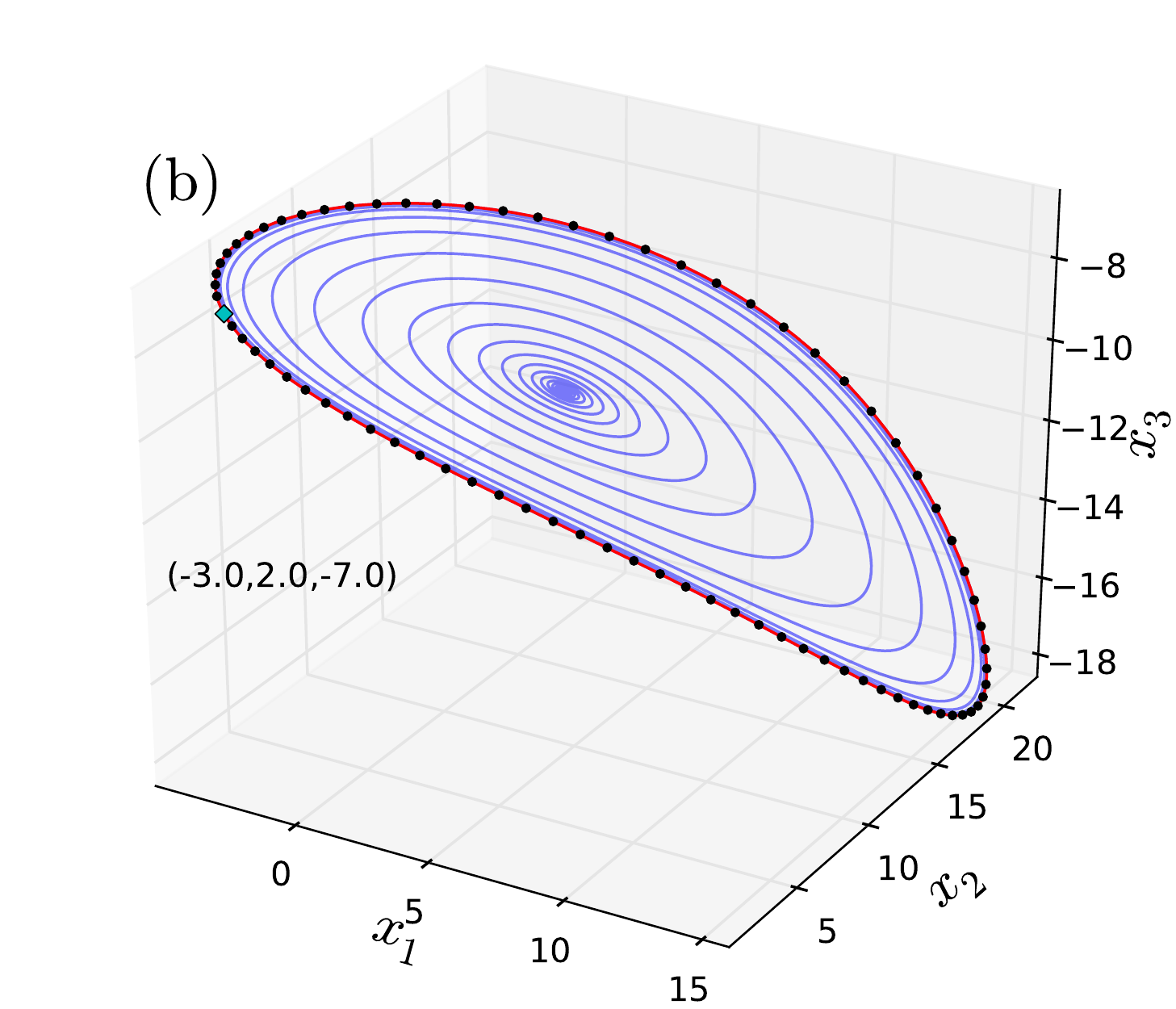}

\includegraphics[width=0.49\textwidth]{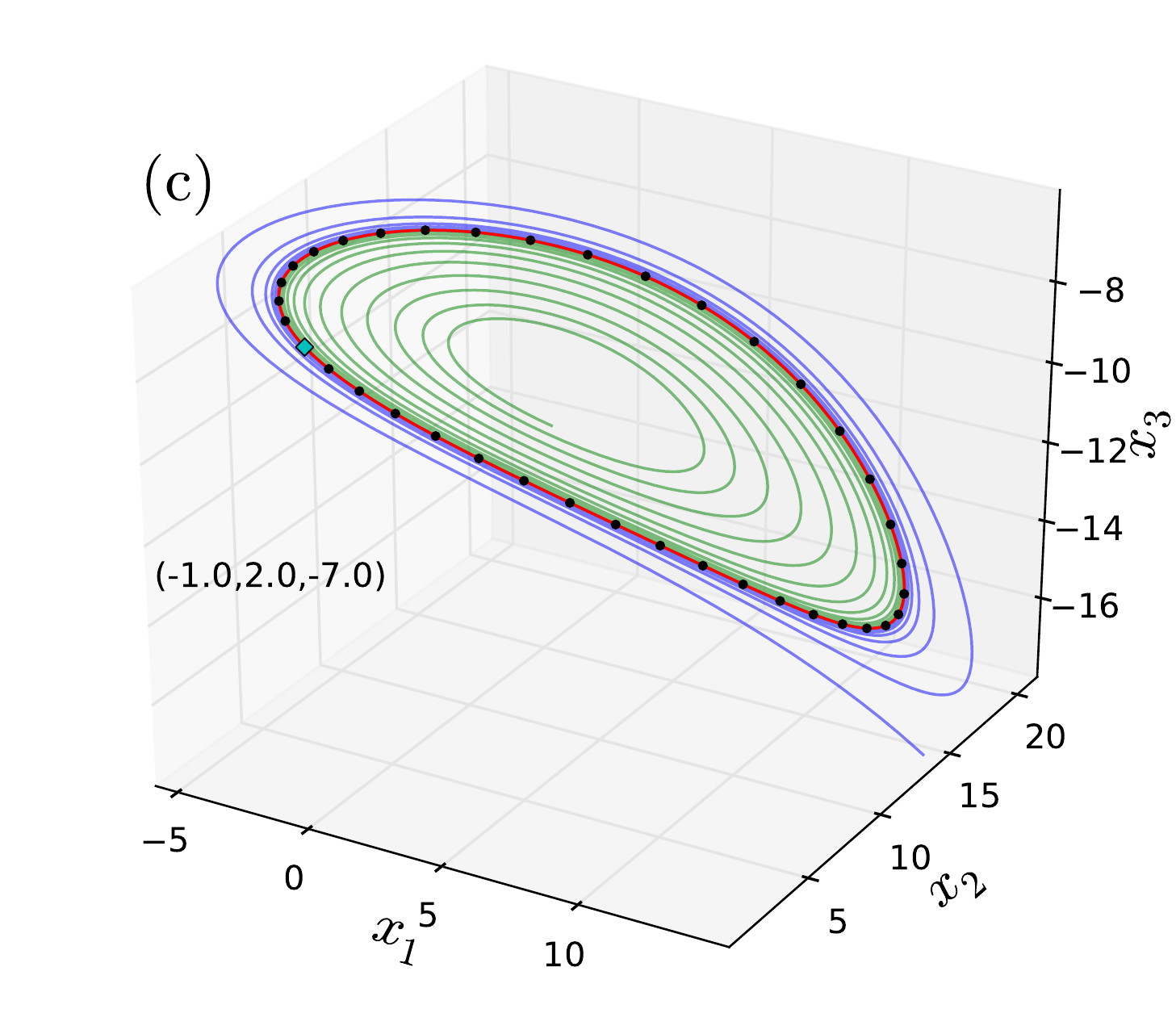}
\includegraphics[width=0.49\textwidth]{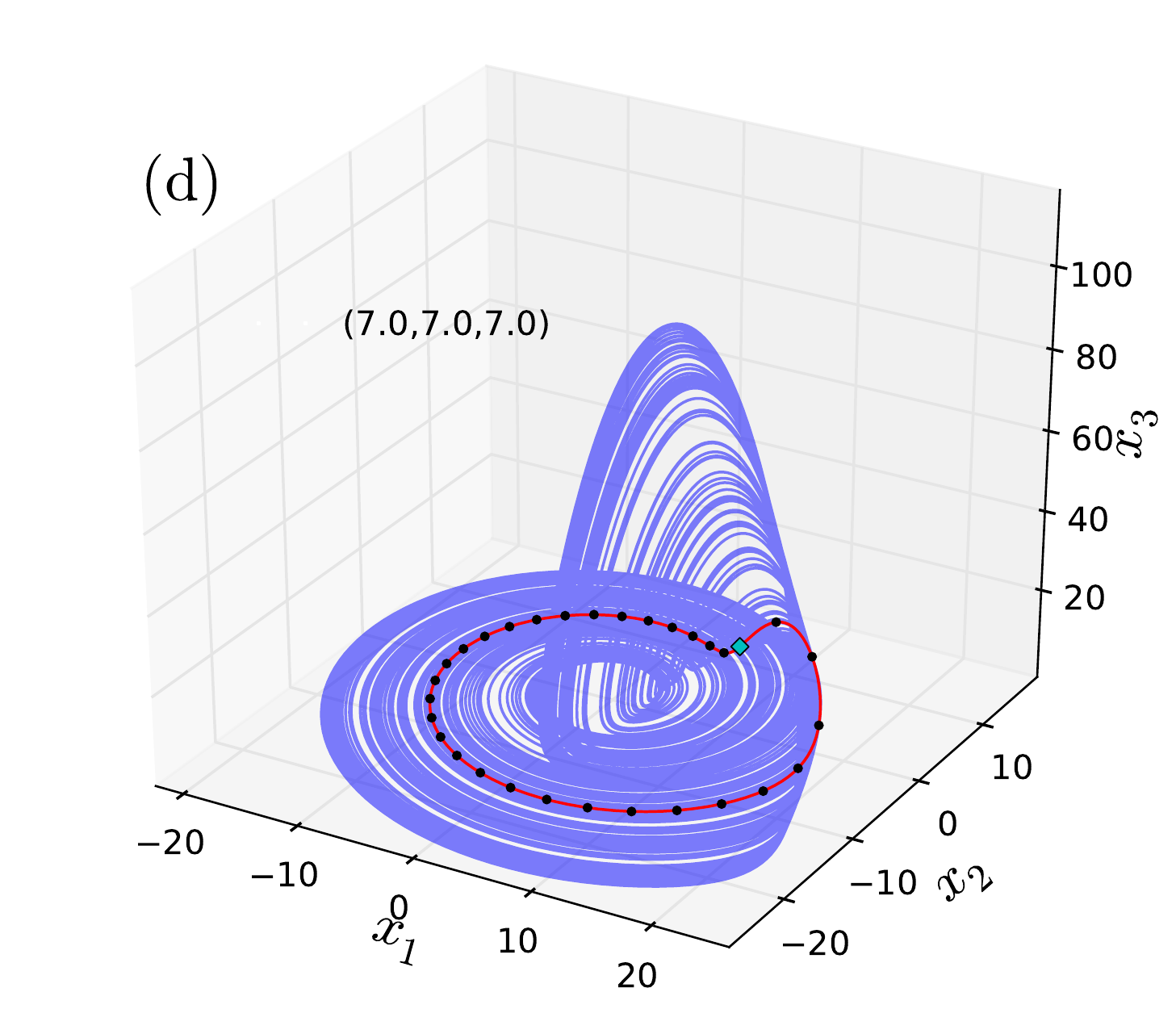}
\caption{(Color online) Different types of paths followed by the
  unstable periodic orbits of the RS after they have decayed: (a)
  Quasi-periodic orbits, (b) and (c) Inward or outward spiral toward
  or away from a fixed point, and (d) Chaotic orbit in basin of
  attraction of a chaotic attractor. The blue and green lines
  spiraling out of and into the fixed point in (c) demonstrate the
  stochastic nature of the decay routes.}
\label{fig4}
\end{figure}

For unstable orbits, numerical errors that occur during the
integration procedure grow larger as the integration proceeds, and
this growth eventually causes the trajectory to move away from the
closed orbit. For the most unstable orbits which we have found, having
$\mu _{\max }\approx 5$, such deviations become visible in the phase
portraits after integration times of $15$-$20T$, where $T$ is the
period of the shortest obtained (period-1) solution. In the least
unstable cases, for which $\mu _{\max }$ is only slightly greater than
one, the deviations become visible after about $80$-$100T$, i.e. after
much  longer integration times, as one would expect. It is also
interesting to note that, after an unstable orbit has decayed, the
trajectory may follow a variety of paths. In some cases we have found
that it  can decay into a quasi-periodic orbit (with essentially an
infinite period), as shown in Figure~\ref{fig4}(a), in other cases it
can spiral inward or outward, either toward or away from a fixed
point, as shown in Figs.~\ref{fig4}(b) and (c), and lastly, it may
enter the basin of attraction of the chaotic attractor, as shown in
Fig.~\ref{fig4}(d). 

The precise parameter values corresponding to all the orbits shown in
Fig.~\ref{fig4} are listed in Table~\ref{tab3} for convenience.

\begin{table}
\begin{tabular}{|rccc|}
\hline Fig. & \multicolumn{1}{|c}{$T$} & \multicolumn{1}{|c}{$R_{\min
  }\left( 10^{-13}\right) $} & \multicolumn{1}{|c|}{$\mu _{\max }$}
\\ \hline\hline 4(a) & \multicolumn{1}{|c}{6.0692872165404559} &
\multicolumn{1}{|c}{0.57} &  \multicolumn{1}{|c|}{5.301613} \\ \hline
4(b) & \multicolumn{1}{|c}{7.6628544529751386} &
\multicolumn{1}{|c}{0.60} &  \multicolumn{1}{|c|}{2.597276} \\ \hline
4(c) & \multicolumn{1}{|c}{7.4670894888859385} &
\multicolumn{1}{|c}{0.23} &  \multicolumn{1}{|c|}{1.844971} \\ \hline
4(d) & \multicolumn{1}{|c}{6.0907519243177850} &
\multicolumn{1}{|c}{1.16} &  \multicolumn{1}{|c|}{4.086460} \\ \hline
&  &  &  \\  &  &  &  \\ \hline Fig. & \multicolumn{1}{|c}{$a$} &
\multicolumn{1}{|c}{$b$} &  \multicolumn{1}{|c|}{$c$} \\ \hline\hline
4(a) & \multicolumn{1}{|c}{0.2591659663922890} & \multicolumn{1}{|c}{
  0.0157490292455600} & \multicolumn{1}{|c|}{28.9026881232841184}
\\ \hline 4(b) & \multicolumn{1}{|c}{-0.5160312363632790} &
\multicolumn{1}{|c}{ -211.2660719529740447} &
\multicolumn{1}{|c|}{27.0217872597939994} \\ \hline 4(c) &
\multicolumn{1}{|c}{-0.4931680147640250} & \multicolumn{1}{|c}{
  -181.9747123841032987} & \multicolumn{1}{|c|}{24.8103827712413185}
\\ \hline 4(d) & \multicolumn{1}{|c}{0.2573147265532380} &
\multicolumn{1}{|c}{ 0.5728277554035760} &
\multicolumn{1}{|c|}{13.6096389284738102} \\ \hline
\end{tabular}
\caption{Optimized periods $T$, minimum residuals $R_{\min}$ and the
  moduli of the largest non-trivial Floquet multipliers $\mu_{\max}$
  corresponding to the unstable periodic orbits shown in
  Fig.~\ref{fig4}, for the optimized  control parameters $a$, $b$ and
  $c$.}
\label{tab3}
\end{table}

We note that in Fig.~\ref{fig4}(c), two spiral trajectories are shown:
the first spiraling inward, the second outward. The difference seen
here, in the way that this orbit decays, occurred as a result of using
two different values for the integration time step. In the first case
$\Delta \tau =1/1024$, and in the second $\Delta \tau =1/2048$. Thus
it is clear that the small  numerical errors that occur during the
integration process can perturb the unstable trajectory in different
ways. This observation emphasizes the stochastic nature of the decay
routes that are depicted in Fig.~\ref{fig4}.

\section{Designing periodic orbits with specific characteristics} \label{sec5}
Unlike the other methods \cite{zho01,li005}, the optimized shooting
method can be used to obtain periodic orbits with very specific
characteristics. To illustrate this feature of the method, we consider
the following  hypothetical example. 

Suppose we are interested in using a Field Programmable Gate Array
(FPGA) implementation of the RS \cite{sad09} to generate a periodic
electrical pulse of a particular amplitude (pulse height). We may
select the coordinate $x_3$,  for this purpose, and assume  that  the
desired amplitude is $(x_3)_{\max}= 318.6$, in the dimensionless units
of Eq. (\ref{eq10}). The problem then is to optimize the parameters
$a$, $b$, $c$, and $T$ (if indeed it is possible), in order to achieve
the desired pulse.

To check the feasibility of this project one can use the optimization
method to search for all stable periodic orbits for which the $x_{3}$
coordinate has the desired maximum.  The only modification that needs
to be made is to the  definition of the residual. In addition to the
components that were  used previously (in Eq.~(\ref{eq10})), extra
components must be added  to the residual in order to define the
additional requirement for the maximum at $x_{3}=318.6$. Thus we must
extend the definition of the residual to
\begin{equation}
\mathbf{R}=(\mathbf{x}\left( 1\right) -\mathbf{x}\left( 0\right)
,b+x_{3}\left( 1\right) \left( x_{1}\left( 1\right) -c\right)
,b+x_{3}\left( 0\right) \left( x_{1}\left( 0\right) -c\right)
)\,\mbox{,}  \label{eq12}
\end{equation}
where the first three components are as before, and the last two
components express the condition for the maximum, i.e. they come from
the requirement $\dot{x}_{3}\left( 1\right) =\dot{x}_{3}\left(0\right)
=0$, with $x_3\left( 0 \right) = 318.6$. The remaining two components
of the initial condition, $x_1 \left( 0 \right) $ and $x_2 \left( 0
\right) $, can of course be chosen arbitrarily in order to set the
phase of the pulse.

\begin{figure}[htp!]
\centering \includegraphics[width=0.49\textwidth]{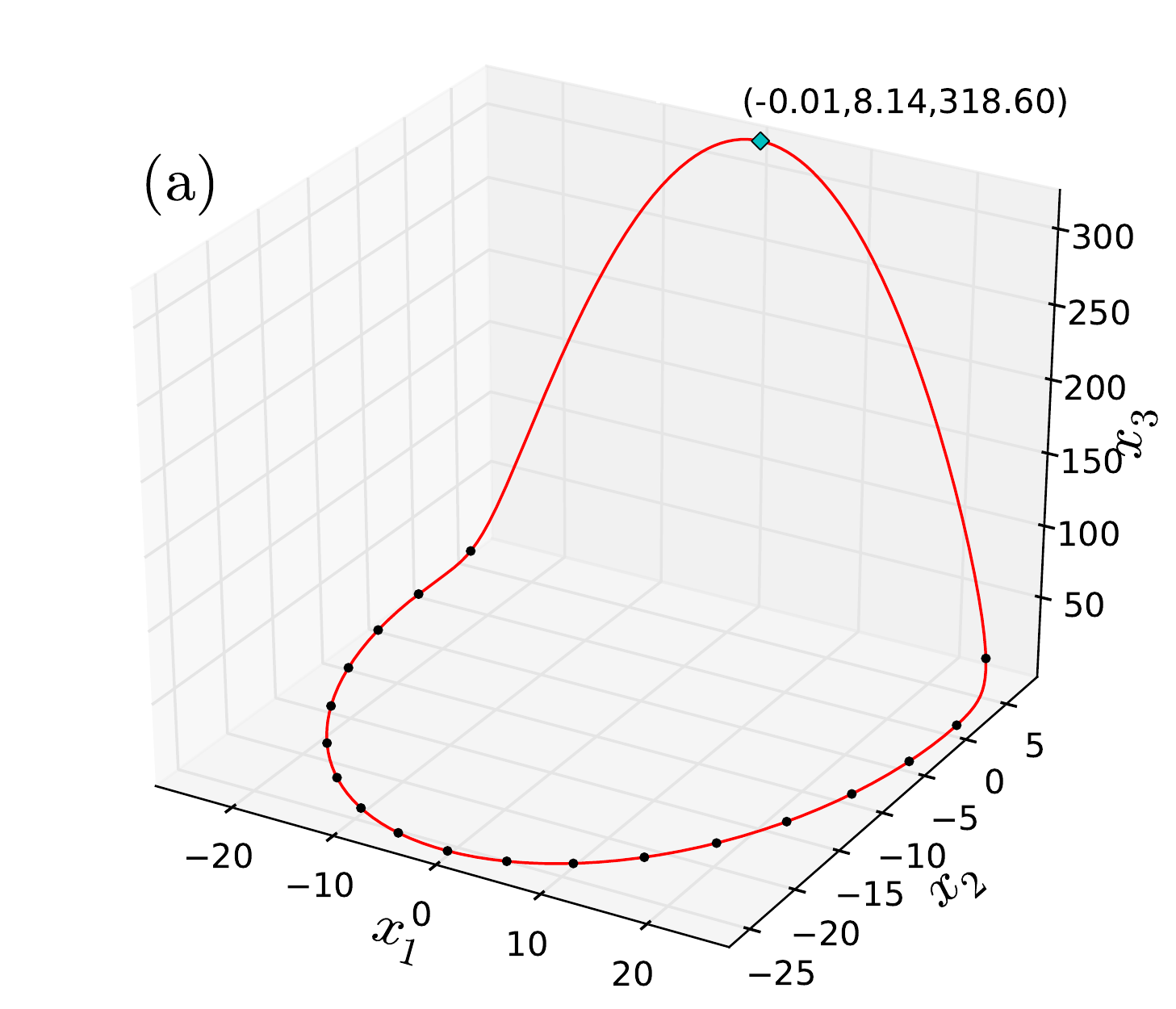}
\includegraphics[width=0.49\textwidth]{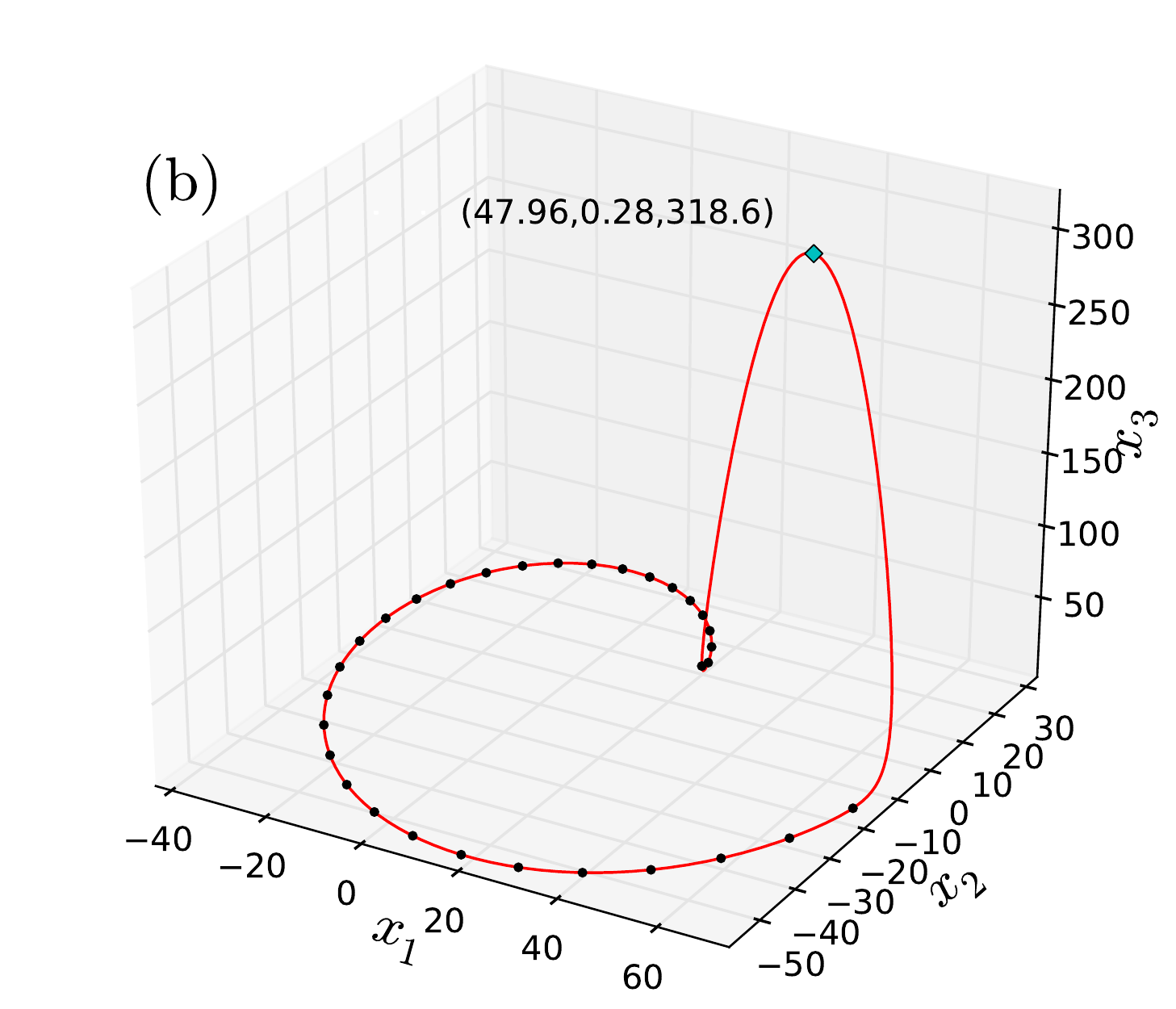}

\includegraphics[width=0.49\textwidth]{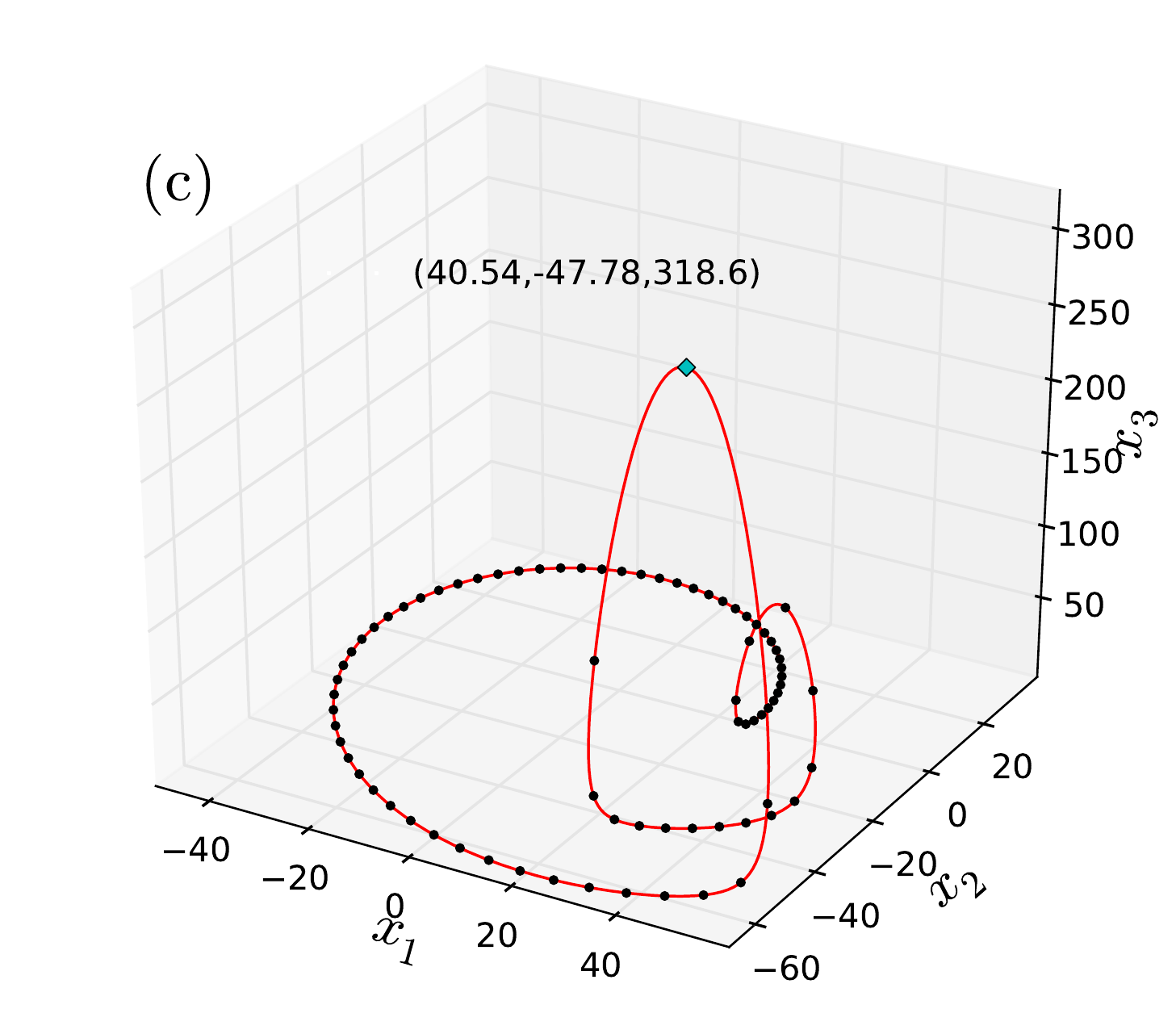}
\includegraphics[width=0.49\textwidth]{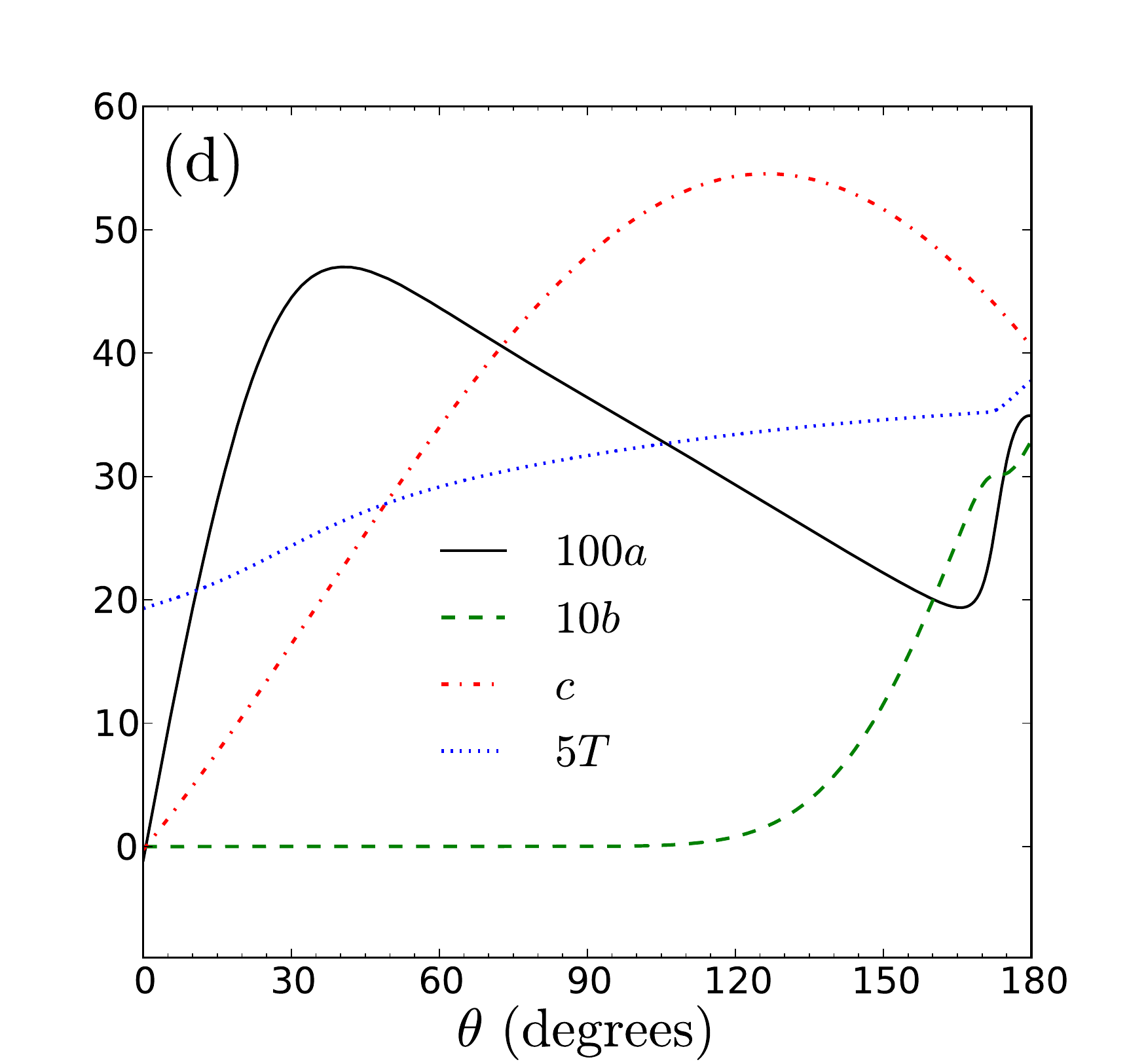}
\caption{(Color online) Illustration of the usefulness of the property
  of the RS discovered in this work. The optimal parameters of the
  periodic orbit in (a) passing exactly through $\left( -0.01\mbox{,
  }8.14\mbox{, }318.60\right) $ are used in a subsequent optimization
  step of an orbit rotated from the previous one by $1$ degree
  relative to an axis parallel to the $x_{3}$-direction and passing
  through the point $\left( 20,-20 \right)$. (b) The orbit obtained
  after $90$ degrees of rotation. (c) After $180$ degrees of rotation.
  (d) The (scaled) optimized parameters as functions of the rotation
  angle.} 
\label{fig5}
\end{figure}

Figure~\ref{fig5}(a) shows the optimized pulse obtained by using the
modified residual (Eq.~(\ref{eq12})), the initial condition
$\left(-0.01, 8.14, 318.60\right)$ , and initial guess for the
parameters:  $a=-0.1, b=0, c=-0.1$, and $T=4$.  
In this case the optimized parameters for the pulse were found after $178$ iterations. In practice the question of how variations in the parameters will affect the phase of
the pulse is also important.

To illustrate this more clearly we re-solve the same problem  after
rotating the initial coordinates $x_1(0)$ and $x_2(0)$ on a circle in
the $x_{1}x_{2}$-plane. Starting from the initial condition
$\left(-0.01, 8.14, 318.60\right)$ we rotate the initial condition by
$1$ degree at a time about an axis parallel to the $x_{3}$-direction and
passing through the point $\left( 20,-20\right) $. For the
optimization of the orbit through the rotated points we choose the the
previously obtained optimized parameters as the initial guess. 

Figure~\ref{fig5}(b) shows the periodic orbit obtained after the first
initial condition, i.e. $\left( -0.01\mbox{, }8.14 \mbox{,
}318.60\right) $, has been rotated clockwise by $90$ degrees about the
axis of rotation. Similarly Fig.~\ref{fig5}(c) shows the orbit after
$180$ degrees of rotation. Figure~\ref{fig5}(d) shows how the
optimized system parameters may be varied continuously in order to
achieve  a smooth rotation in the phase of the pulse. In this example
we have allowed the period $T$ of the pulse to vary while the phase is
being rotated. However, there are many other possibilities afforded by
the present method. For example, one  can also rotate the phase of the
pulse for a fixed period. 

It is interesting to note that, as one perhaps may have anticipated,
for the second and subsequent optimizations in the above example  the
number of iterations required to reach the set tolerance of $R_{\min
}<10^{-14}$ become considerably fewer than the initial number of
$178$. The subsequent  optimizations only require $20$ to $35$
iterations to converge.

\section{Periodic solutions of a coupled R\"{o}ssler system}
To test the optimization method on a system that  exhibits
high-dimensional chaos,  we have also applied it to a six-dimensional
(symmetrically) coupled  RS \cite{ros96}, using for comparison the
same parameter values that were considered in Ref.~\cite{zho01}. The
study of the dynamics of identical coupled nonlinear chaotic flows,
such as two coupled RS \cite{ras96,yan01,pra13}, has provided insights
into the chaotic behavior of higher dimensional systems. These systems
exhibit hyperchaos, which differs from ordinary chaos in that there is
more than one positive Lyapunov exponent, and hence more than one
direction in phase space in which the chaotic attractor  expands
\cite{ras96,pra13}. 

In the notation of Ref.~\cite{zho01}, the system equations are given
by
\begin{eqnarray}
\dot{x}_{1} &=&-w_{1}x_{2}-x_{3}+c\left( x_{4}-x_{1}\right) \nonumber
\\ \dot{x}_{2} &=&w_{1}x_{1}+0.15x_{2}  \nonumber \\ \dot{x}_{3}
&=&0.2+x_{3}\left( x_{1}-3.5\right)   \label{eq14} \\ \dot{x}_{4}
&=&-w_{2}x_{5}-x_{6}+c\left( x_{1}-x_{4}\right) \nonumber
\\ \dot{x}_{5} &=&w_{2}x_{4}+0.15x_{5}  \nonumber \\ \dot{x}_{6}
&=&0.2+x_{6}\left( x_{4}-3.5\right)   \nonumber
\end{eqnarray}
where $w_{1}=1.03$, $w_{2}=0.97$ and the coupling constant $c=0.13$.

To test our method on this six-dimensional system, we started from the
initial condition $(1,0,1,0,1,0)$ and an initial guess, $T=5$, for the
period.
\begin{figure}[htp!]
\centering \includegraphics[width=0.49\textwidth]{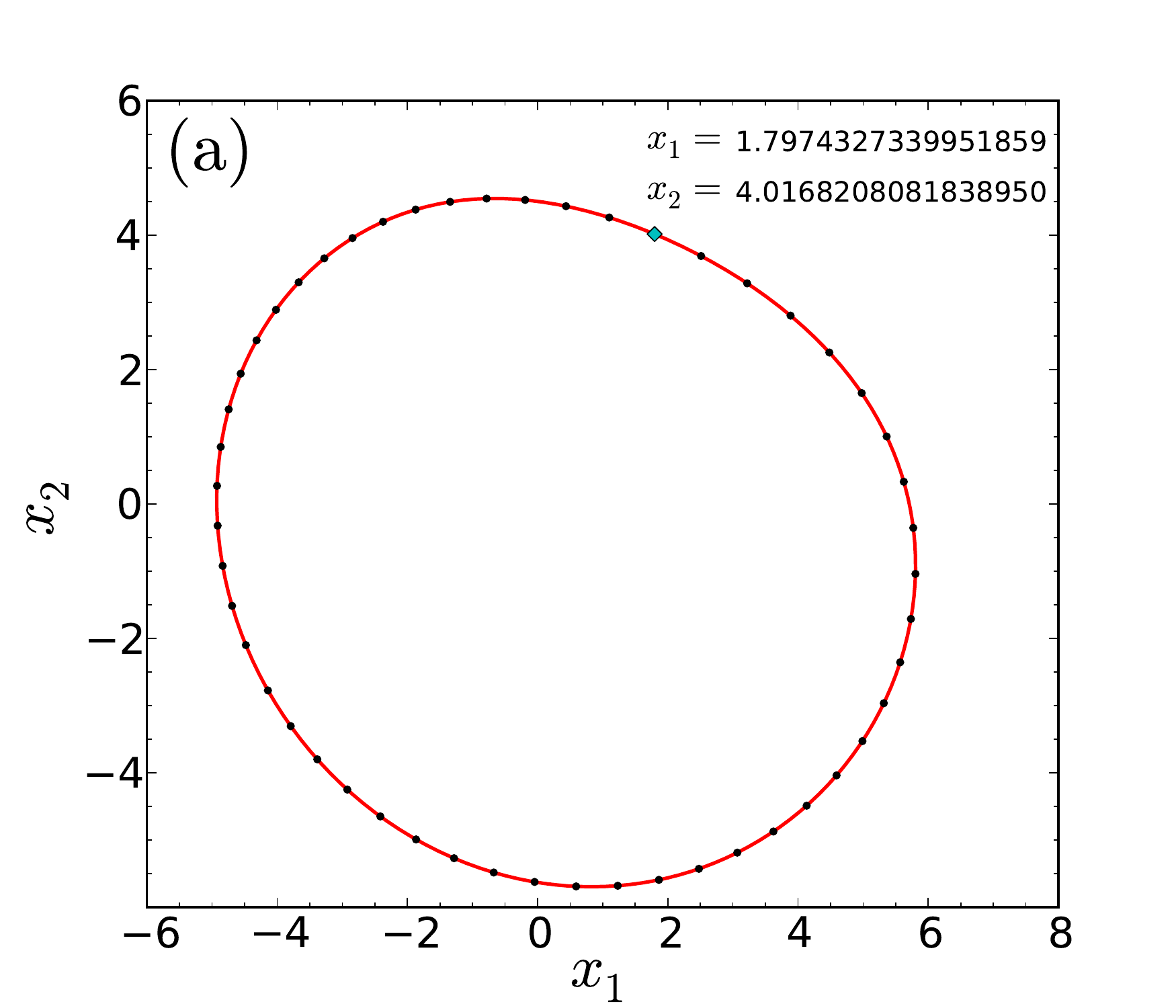}
\includegraphics[width=0.49\textwidth]{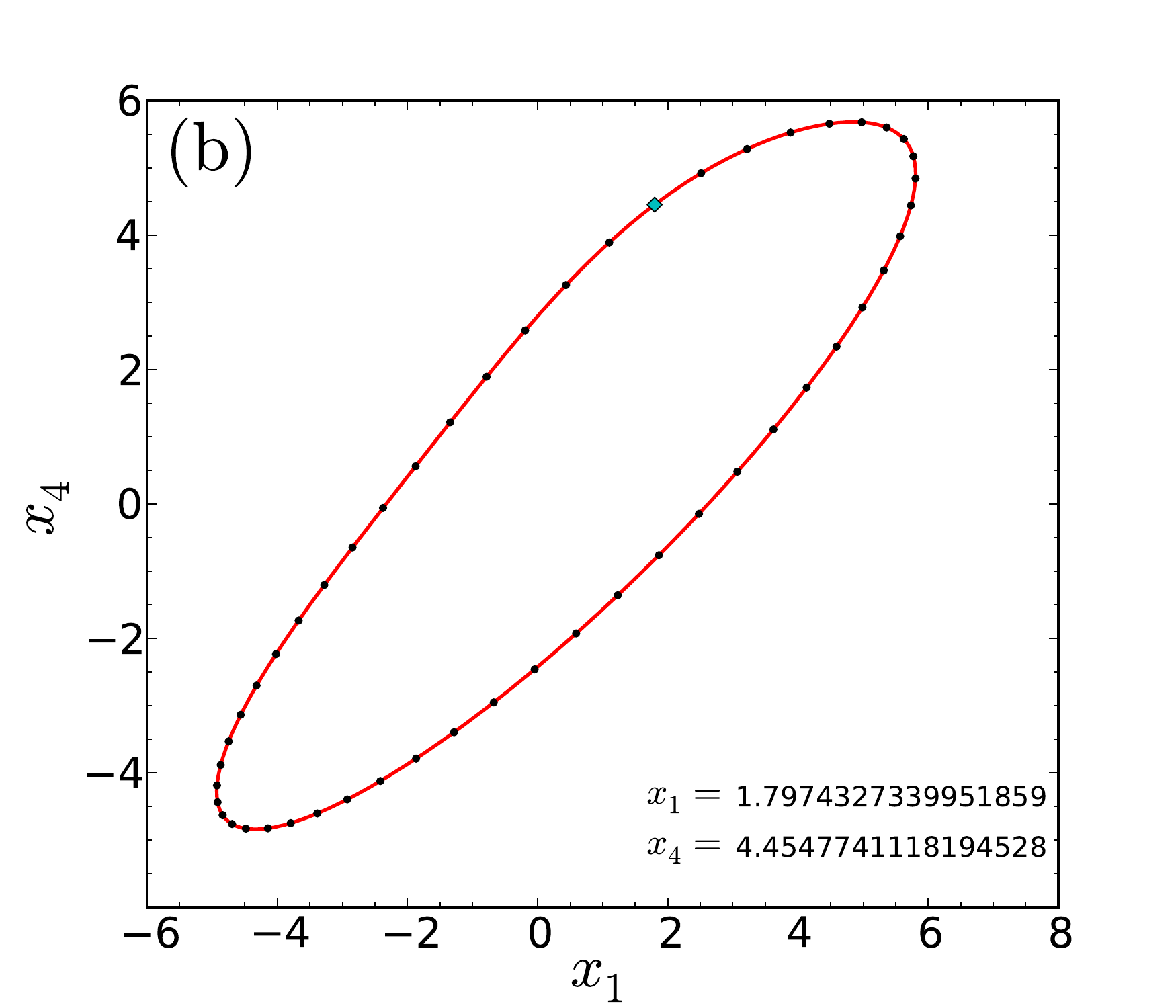}

\includegraphics[width=0.49\textwidth]{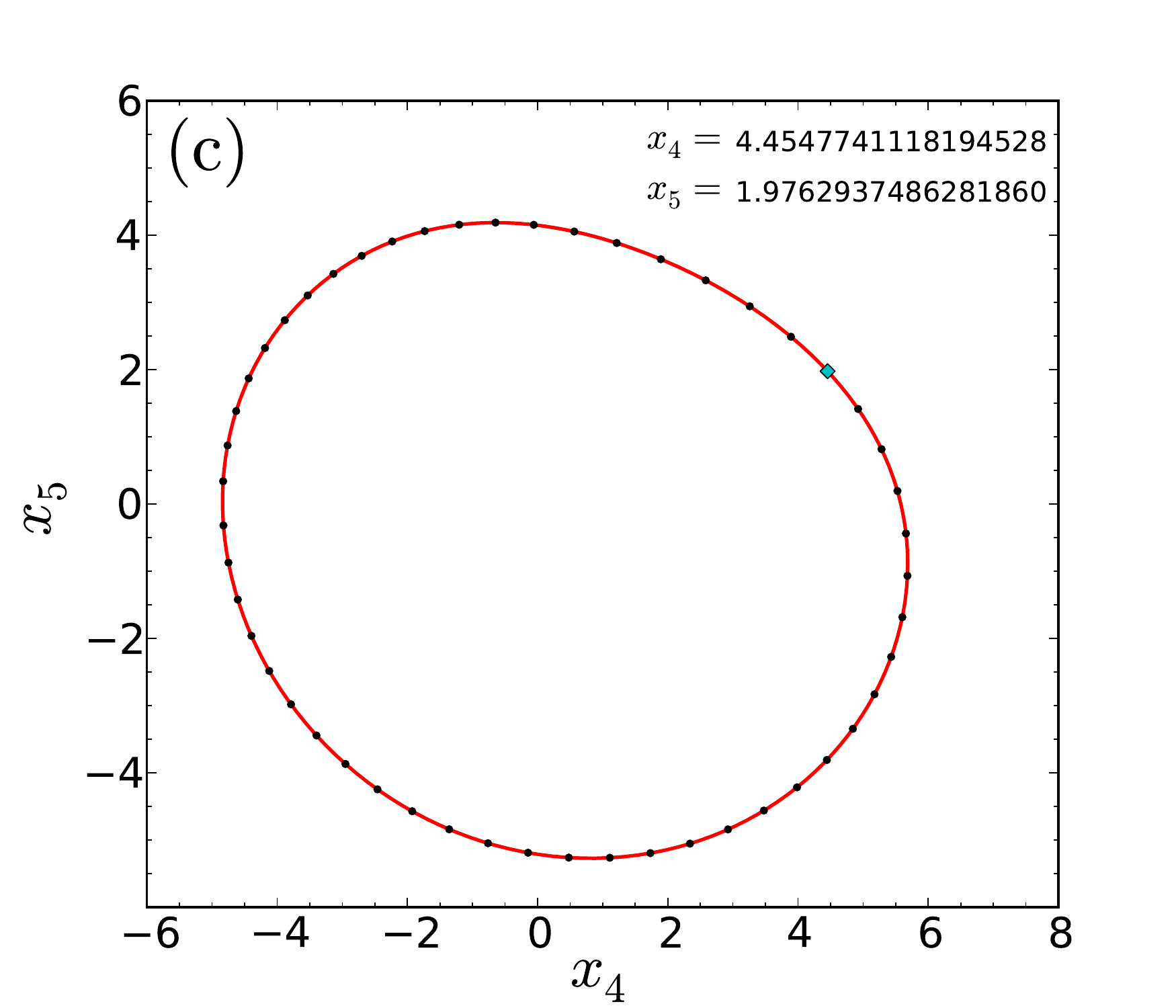}
\includegraphics[width=0.49\textwidth]{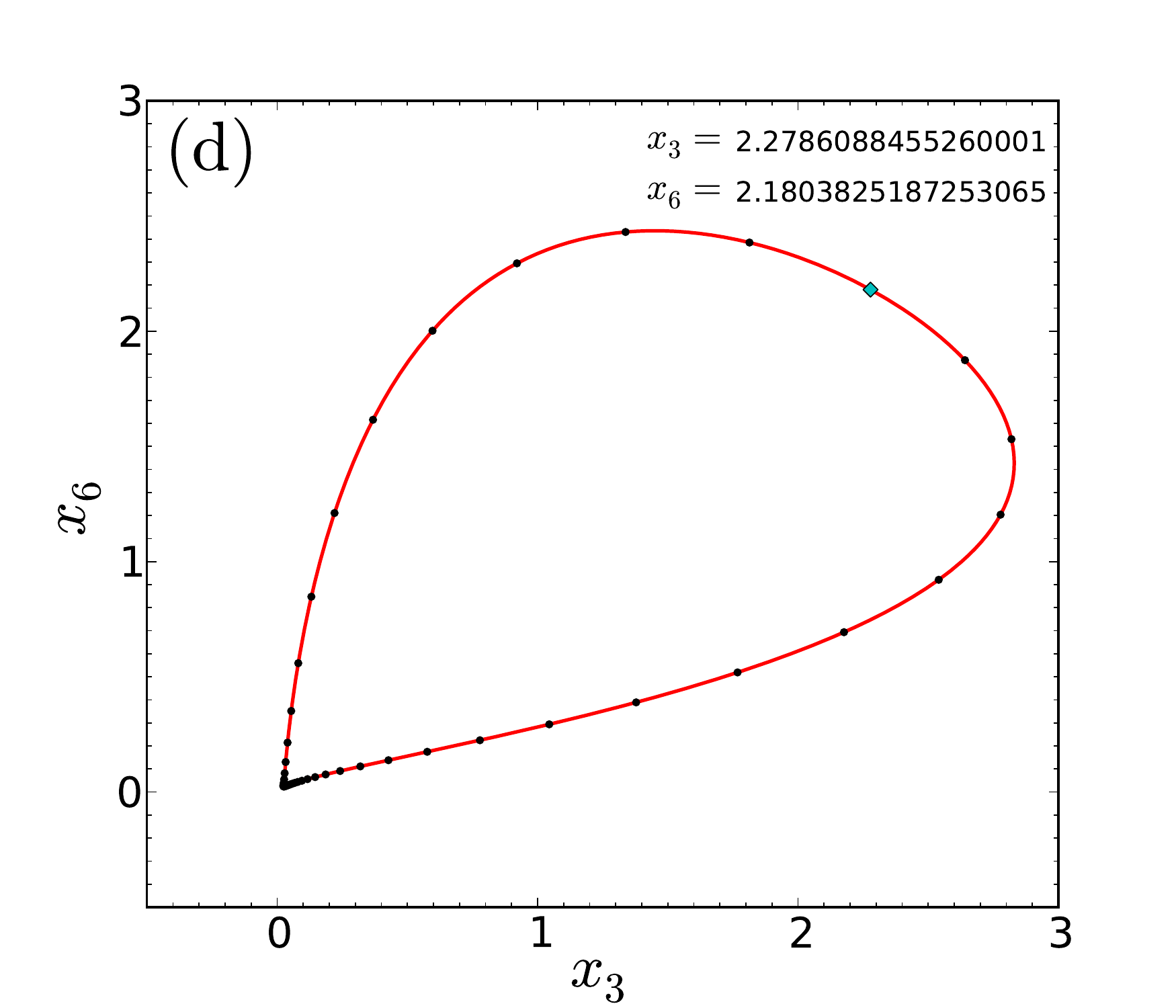}
\caption{(Color online) Phase portraits of the periodic orbit found by
  the optimized shooting method, in agreement with the results of
  Ref.~\cite{zho01}. The precise coordinates of one point on the
  orbit, indicated by a blue diamond marker, are listed in each
  figure.}
\label{fig6}
\end{figure}
The optimized period of $T=5.9773863584207021$ was found for the
period-1 orbit, with an error in the residual $\sim
10^{-14}$. Figure~\ref{fig6} shows four projections of the phase
portrait for the calculated periodic orbit, together with one point on
the orbit. Figures~\ref{fig6}(a)-(c) are the same as Figs. 2-4 in the
paper by Zhou \emph{et al}.~\cite{zho01} The magnitude of the  largest
non-trivial Floquet multiplier for this orbit was found to be
$\mu_{\max}=0.649768$, indicating that the orbit is stable. For other
values of the control parameters the system was found to exhibit
quasi-periodic behaviour, interrupted by periodic windows in which
frequency-locking occurs, in agreement with Ref.~\cite{ras96}.

\section{Periodic solution of a flexible rotor-bearing system}
To demonstrate that our method is equally applicable to non-autonomous
systems, we model the Jeffcott flexible rotor-bearing system that was
tested by Li and Xu~\cite{li005}. This non-autonomous system  also
exhibits high-dimensional chaos. In the notation of
Ref.~\cite{li005}, the system equation is given by
\begin{eqnarray}
\dot{x}_{i} &=&x_{i+4}\mbox{ \ \ (}i=1,2,3,4\mbox{) }  \nonumber
\\ \dot{x}_{5} &=&-\frac{\sigma }{\omega }f_{x}+\frac{d}{\omega
}\left( x_{7}-x_{5}\right) +\frac{k}{\omega ^{2}}\left(
x_{3}-x_{1}\right)   \nonumber \\ \dot{x}_{6} &=&-\frac{\sigma
}{\omega }f_{y}+\frac{d}{\omega }\left( x_{8}-x_{6}\right)
+\frac{k}{\omega ^{2}}\left( x_{4}-x_{2}\right) -\frac{1}{ \omega
  ^{2}}  \label{eq15} \\ \dot{x}_{7} &=&\beta \cos \tau
+\frac{d}{\gamma \omega }\left( x_{5}-x_{7}\right) +\frac{k}{\gamma
  \omega ^{2}}\left( x_{1}-x_{3}\right)   \nonumber \\ \dot{x}_{8}
&=&\beta \sin \tau +\frac{d}{\gamma \omega }\left( x_{6}-x_{8}\right)
+\frac{k}{\gamma \omega ^{2}}\left( x_{2}-x_{4}\right) -
\frac{1}{\omega ^{2}}  \nonumber
\end{eqnarray}
where $\sigma =0.843$, $k=5893.9$, $d=7.677$, $\gamma =16$, $\beta
=0.14328$ and $\omega $ is the variable parameter which is related to
the angular frequency at which the rotor is driven. In the last
equation the overdot indicates the total time derivative with respect
to the dimensionless time $\tau$ and the components of the nonlinear
oil film force are
\begin{equation}
f_x=-p_r\cos\psi - p_t\sin\psi \mbox{ and \ }
f_y=-p_r\sin\psi + p_t\cos\psi \mbox{, }\label{eq16}
\end{equation}
where
\begin{eqnarray*}
p_r & = & \frac{6\varepsilon^2 (1-2\dot{\psi})}{(1-\varepsilon^2)(2+\varepsilon^2)}
+ \frac{3\dot{\varepsilon}}{(1-\varepsilon^2)^{3/2}}\left( \pi - \frac{16}{\pi(2+\varepsilon^2)}\right) \mbox{ and} \\
p_t & = & \frac{3\pi\varepsilon (1-2\dot{\psi})}{(1-\varepsilon^2)^{1/2}(2+\varepsilon^2)} + \frac{12\varepsilon\dot{\varepsilon}}{(1-\varepsilon^2)(2+\varepsilon^2)}
\end{eqnarray*}
are the radial and tangential forces on the oil film. (See
Refs.~\cite{cho13,sha90} for details.) We note here that it was
extremely difficult to reconstruct the full set of equations from the
information provided in the paper by Li and Xu~\cite{li005}, firstly
because of two typographical errors in their equations, and secondly
because their equations were incomplete, with the only references
provided to unavailable  papers in Chinese. In view of these
difficulties we have corrected the  typographical errors and provided
(with the help of Refs.~\cite{cho13,sha90}  and the references
therein) the missing transformation equations, 
\begin{equation}
x_{1}=-\varepsilon \cos \psi \mbox{ and \ }  x_{2}=-\varepsilon \sin
\psi  \mbox{, }  \label{eq17}
\end{equation}
that allow one to express the polar coordinates, $\varepsilon$ and
$\psi$, and  their total time derivatives, in terms of the dynamical
variables $x_{1}$ and $x_{2}$.

\begin{figure}[htp!]
\centering \includegraphics[width=0.49\textwidth]{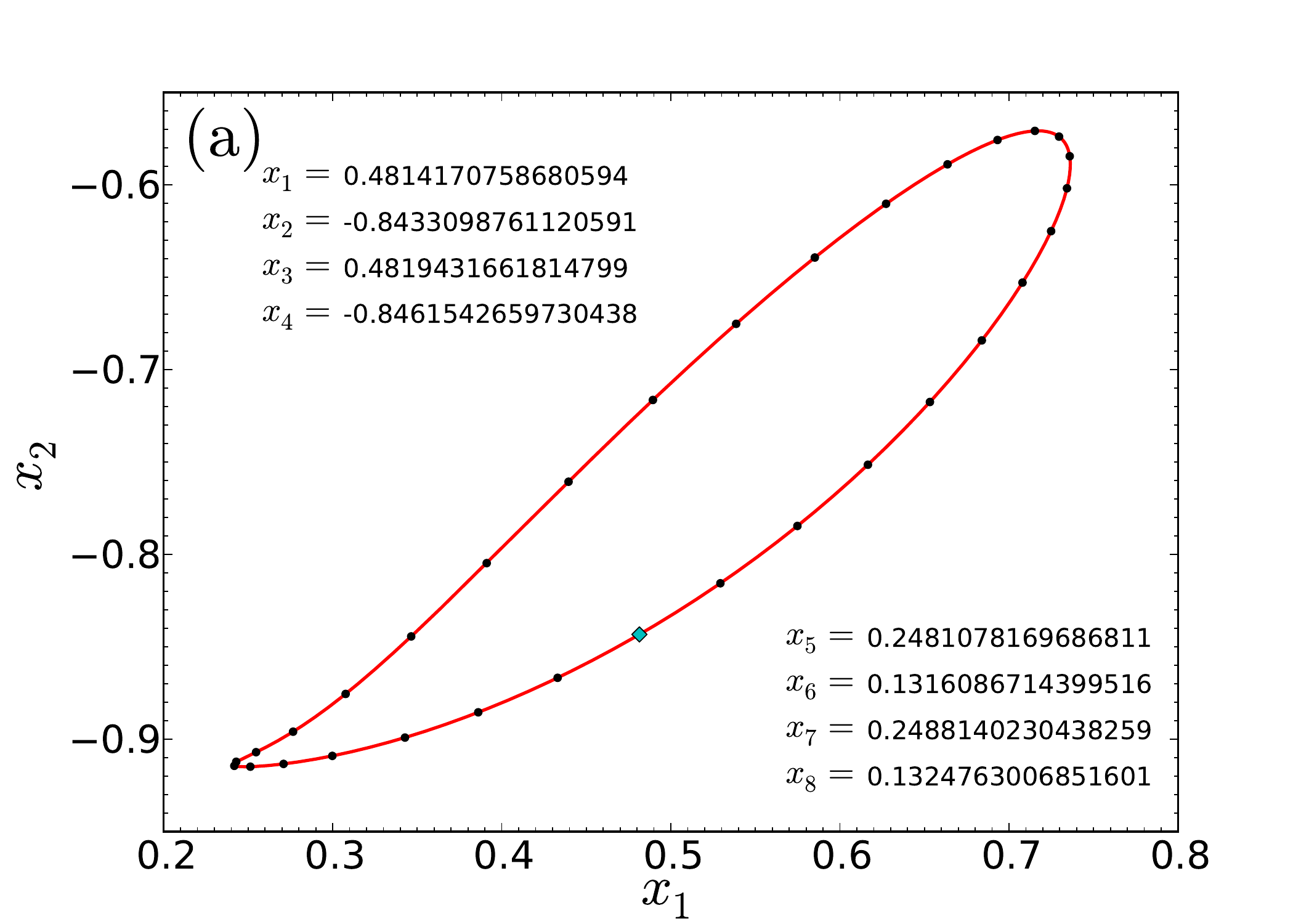}
\includegraphics[width=0.49\textwidth]{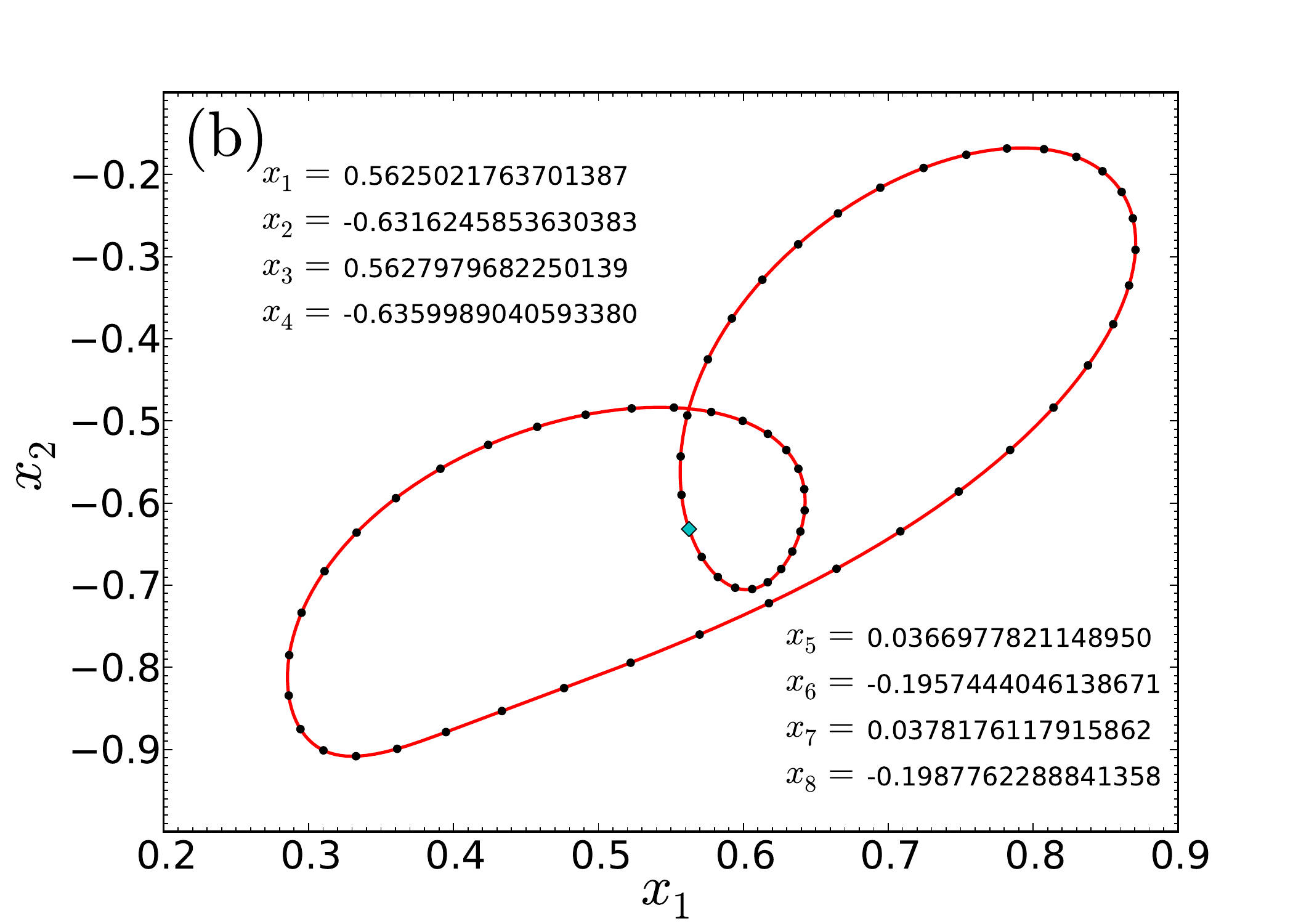}

\caption{(Color online) Phase portraits reproduced from
  Ref.~\cite{li005} for the Jeffcott rotor-bearing system by the
  numerical method described in this paper. (a) The period-1 solution
  of Eq.~(\ref{eq15}) obtained from the initial conditions shown when
  $\omega = 1.2 $. (b) The period-2 solution obtained when $\omega =
  2.3$ and the starting conditions shown are used.}
\label{fig7}
\end{figure}

Figure~\ref{fig7} shows good agreement for the quantities that were
plotted in Figs.~6 and 7 of Ref.~\cite{li005}. Starting from the same
initial conditions that were used in Ref.~\cite{li005}, we calculated
the period of the period-1 orbit ($\omega=1.2$) to be
$T=6.2831853071795498$. The phase portrait for the corresponding orbit
is shown in Fig.~\ref{fig7}(a), together with the coordinates of one
point on the orbit. For the period-2 orbit ($\omega=2.3$), shown in
Fig.~\ref{fig7}(b), our method  produces an optimized period of  $T=
12.5663706143591707$. The calculated  periods agree with the
theoretically expected values of $2\pi$ ($4\pi$) to an impressive
thirteen (fourteen) decimal places.

\section{Conclusion}
We have developed an optimized shooting method for finding the
periodic solutions of both autonomous and non-autonomous nonlinear
systems of differential equations. The method essentially re-casts the
problem of finding periodic orbits in a form that is suitable for
applying multi-dimensional optimization. In the present work we have 
made use of Levenberg-Marquardt optimization (LMO) for this purpose. LMO 
is widely regarded  as being one of the most efficient methods of optimization 
for medium-sized problems, i.e. for those with up to a few hundred weights. 

Since LMO is already available in the standard libraries for the most
important  scientific programming languages, such as Python, Fortran,
Matlab, C and C++, our present method for finding periodic orbits  is
relatively easy to implement and does not require additional
resources.  In the present work we have provided a simple
implementation of the method in  the Python programming language. Even
though we have only explored a few possible ways of defining the
residual (error vector) that is required for  the method, we would
like to emphasize that the true versatility of the method ultimately
depends on the user's ability to define  the residual
appropriately. Nevertheless we hope that the present  examples have
served to illustrate the basic idea behind the method and that they
will in future stimulate the creative use of the optimized shooting method
for finding periodic orbits in nonlinear dynamical systems. 

\begin{acknowledgements}
The authors would like to thank M.R. Kolahchi, G. Qi,
J.R. Ruiz-Femenia and J.A. Caballero-Suarez for helpful discussions
about this work.
\end{acknowledgements}

\appendix
\section*{Appendix A: Example of computer implementation}
The following code, written in the Python programming
language \cite{chu07}, sets up and minimizes the residual vector
$\mathbf{R}$, given by Eq.~(\ref{eq10}), excluding the $x_{3}$-coordinate 
from the minimization.  Although there is an integration scheme
which is specially adapted to orbital problems such as 
these \cite{ana05}, the present example makes use of a fifth-order
Runge-Kutta scheme that already produces excellent results.  In the
code below the function \verb|f()| returns the derivatives for the
R\"{o}ssler system. The function \verb|ef()| returns the residual by 
integrating the system from $ \tau =0$ to $\tau =1+p\Delta \tau $, through calls to \verb|integrate()|. In the present example,
$p=2$, and the step size is set to $\Delta \tau =1/1024$. The code
given below is designed to work with step sizes of the form $1/2^{\ell}$,
where $\ell$ is an appropriately chosen positive integer. The function
\verb|leastsq()|, which is imported from the module
\verb|scipy.optimize| \cite{lan04}, uses LMO to minimize
the residual defined in the function \verb|ef()|. The function 
\verb|leastsq()| is called from within the \verb|main()| function, which
sets up the initial parameters and quantities to be
optimized. Notice that in this example, only the three quantities
$x_{1}$, $x_{2}$ and $T$ are passed to \verb|leastsq()| for
optimization, via the vector \verb|v0|.
\begin{verbatim}
from scipy import zeros, concatenate, sqrt, dot
from scipy.optimize import leastsq
 
def f(t,x,T,a):
    """
    Rossler system written in the form of Eq. (7)
    """
    xd = zeros(len(x),'d')
    xd[0] = T*(-x[1]-x[2])
    xd[1] = T*(x[0]+a[0]*x[1])
    xd[2] = T*(a[1]+x[2]*x[0]-a[2]*x[2])
    return xd
 
def integrate(t,x,func,h,w,a):
    """
    5th-order Runge-Kutta integration scheme. Input:
    t - initial time
    x - vector of initial conditions at initial time t
    h - integration time step, w - period
    a - additional parameters
    """
    k1=h*func(t,x,w,a)
    k2=h*func(t+0.5*h,x+0.5*k1,w,a)
    k3=h*func(t+0.5*h,x+(3.0*k1+k2)/16.0,w,a)
    k4=h*func(t+h,x+0.5*k3,w,a)
    k5=h*func(t+h,x+(-3.0*k2+6.0*k3+9.0*k4)/16.0,w,a)
    k6=h*func(t+h,x+(k1+4.0*k2+6.0*k3-12.0*k4+8.0*k5)/7.0,w,a)
    xp = x + (7.0*k1+32.0*k3+12.0*k4+32.0*k5+7.0*k6)/90.0
    return xp
 
def ef(v,x,func,dt,a,p):
    """
    Residual (error vector). Input:
    v - vector containing the quantities to be optimized
    x - vector of initial conditions
    func - function, dt - integration time step
    a - additional parameters
    p - controls length of error vector
    """
    j = int(2.0/dt)
    vv = zeros((j,len(x)),'d')
    vv[0,0:2] = v[0:2]       # set initial condition
    vv[0,2] = x[2]
    T = v[2]                 # set period
    i = 0
    while i < j/2+p:
        t = i*dt
        vv[i+1,:] = integrate(t,vv[i,:],func,dt,T,a)
        i = i+1
    er = vv[j/2,:]-vv[0,:]   # creates residual error vector
    for i in range(1,p):     # of appropriate length
        er = concatenate((er,vv[j/2+i,:]-vv[i,:]))
    print 'Error:', sqrt(dot(er,er))
    return er
 
def main():
    a0 = zeros(3,'d')        # predetermined system parameters
    a0[0] = 0.15; a0[1] = 0.2; a0[2] = 3.5
    x0 = zeros(3,'d')        # initial conditions (N=3)
    x0[0] = 2.7002161609; x0[1] = 3.4723025491; x0[2] = 3.0
    v0 = zeros(3,'d')        # quantities for optimization
    v0[0:2] = x0[0:2]
    v0[2] =   5.92030065     # initial guess for period
    p = 2                    # length of residual is pN
    h = 1.0/1024.0           # integration time step
    #                        # LM optimization
    v, succ = leastsq(ef,v0,args=(x0,f,h,a0,p),ftol=1e-12,maxfev=200)
    err = ef(v,x0,f,h,a0,p)  # error estimation
    es = sqrt(dot(err,err))
    #
    fout = open('fig1ab.dat','w') # for file output
    u0 = (v[0],v[1],x0[2],v[2],es/1e-13)
    print ('%20.16f %20.16f %20.16f %20.16f %6.2f' % u0)
    print >> fout,('%20.16f %20.16f %20.16f %20.16f %6.2f' % u0)
    fout.close()
 
main()
\end{verbatim}
The above code executes in 3.78 CPU seconds on an Intel 3.0 GHz Xeon
processor and requires a maximum memory (RAM) of 2MB, with at least 30MB of
additional swap space. The output of the code is written to screen as well
as to the file called \verb|fig1ab.dat|. After execution the file will
contain the following numbers:
\begin{verbatim}
2.6286556703142154 3.5094562051716300 3.0000000000000000 
5.9203402481939138 0.21
\end{verbatim}
Here the quantities are: the optimized initial point on the periodic orbit,
the corresponding period, and the magnitude of the final value of the
residual, divided by $10^{-13}$.
\bibliographystyle{spphys}
\bibliography{dednam}
\end{document}